\documentclass[iop]{emulateapj}
\usepackage{graphicx}
\usepackage{booktabs}
\usepackage{amsmath}
\usepackage{amssymb}
\usepackage{amsthm}
\usepackage[backref,breaklinks,colorlinks,citecolor=blue]{hyperref}
\usepackage[caption=false]{subfig}
\usepackage{color}
 \usepackage[capitalise]{cleveref}

\usepackage{listings}
\newcommand{\nver}{\hat{\mathbf{n}}}

\def\lsim{\,\lower2truept\hbox{${<\atop\hbox{\raise4truept\hbox{$\sim$}}}$}\,}
\def\gsim{\,\lower2truept\hbox{${> \atop\hbox{\raise4truept\hbox{$\sim$}}}$}\,}

\begin{document}
\title{Toward a tomographic analysis of the cross-correlation between \textit{Planck} CMB lensing and H-ATLAS galaxies}

%%%%%%%%%%%%%
%%      AUTHORS       %%
%%%%%%%%%%%%%

\author{F. Bianchini\altaffilmark{1,4,5}, A. Lapi\altaffilmark{2,1,4,5}, M. Calabrese\altaffilmark{1,12} P. Bielewicz\altaffilmark{1,6},
J. Gonzalez-Nuevo\altaffilmark{3}, C. Baccigalupi\altaffilmark{1,4},  L. Danese\altaffilmark{1}, G. de Zotti\altaffilmark{7,1}, N. Bourne\altaffilmark{8}, A. Cooray\altaffilmark{11}, L. Dunne\altaffilmark{9,8,10}, S. Eales\altaffilmark{9}, E. Valiante\altaffilmark{9}}

\altaffiltext{1}{Astrophysics Sector, SISSA, Via Bonomea 265, I-34136 Trieste, Italy; fbianchini@sissa.it}
\altaffiltext{2}{Dipartimento di Fisica, Universit\`a ``Tor Vergata'', Via della Ricerca Scientifica 1, I-00133 Roma, Italy}
\altaffiltext{3}{Departamento de F\'{i}sica, Universidad de Oviedo, C. Calvo Sotelo s/n, 33007 Oviedo, Spain}
\altaffiltext{4}{INFN - Sezione di Trieste, Via Valerio 2, I-34127 Trieste, Italy}
\altaffiltext{5}{INAF - Osservatorio Astronomico di Trieste, via Tiepolo 11, 34131, Trieste, Italy}
\altaffiltext{6}{Nicolaus Copernicus Astronomical Center, Bartycka 18, 00-716 Warsaw, Poland}
\altaffiltext{7}{INAF - Osservatorio Astronomico di Padova, Vicolo dell'Osservatorio 5, I-35122 Padova, Italy}
\altaffiltext{8}{Scottish Universities Physics Alliance (SUPA), Institute for Astronomy, University of Edinburgh, Royal Observatory, Edinburgh EH9 3HJ }
\altaffiltext{9}{School of Physics and Astronomy, Cardiff University, Queens Buildings, The Parade, Cardiff CF24 3AA, UK}
\altaffiltext{10}{Department of Physics and Astronomy, University of Canterbury, Private Bag 4800, Christchurch, 8140, New Zealand}
\altaffiltext{11}{Department of Physics and Astronomy, University of California, Irvine CA 92697, USA}
\altaffiltext{12}{INAF, Osservatorio Astronomico di Brera, via E. Bianchi 46, I-23807  (LC), Italy}

\email{fbianchini@sissa.it}

%%%%%%%%%%%%%
%%      ABSTRACT      %%
%%%%%%%%%%%%%

\begin{abstract}
We present an improved and extended analysis of the cross-correlation between the map of the Cosmic Microwave Background (CMB) lensing potential derived from the \emph{Planck} mission data and the high-redshift galaxies detected by the \emph{Herschel} Astrophysical Terahertz Large Area Survey (H-ATLAS) in the photometric redshift range $z_{\rm ph} \ge 1.5$. We compare the results based on the 2013 and 2015 \textit{Planck} datasets, and investigate the impact of different selections of the H-ATLAS  galaxy samples. Significant improvements over our previous analysis have been achieved thanks to the higher signal-to-noise ratio of the new CMB lensing map recently released by the \textit{Planck} collaboration. The effective galaxy bias parameter, $b$, for the full galaxy sample, derived from a joint analysis of the cross-power spectrum and of the galaxy auto-power spectrum  is found to be $b = 3.54^{+0.15}_{-0.14}$. Furthermore, a first tomographic analysis of the cross-correlation signal is implemented, by splitting the galaxy sample into two redshift intervals: $1.5 \le z_{\rm ph} < 2.1$ and $z_{\rm ph}\ge 2.1$. A statistically significant signal was found for both bins, indicating a substantial increase with redshift of the bias parameter: $b=2.89\pm0.23$ for the lower and $b=4.75^{+0.24}_{-0.25}$ for the higher redshift bin. Consistently with our previous analysis we find that the amplitude of the cross correlation signal is a factor of $1.45^{+0.14}_{-0.13}$ higher than expected from the standard $\Lambda$CDM model  for the assumed redshift distribution. The robustness of our results against possible systematic effects has been extensively discussed although the tension is mitigated by passing from 4 to 3$\sigma$.

%% Context.
%% Aims.
%% Results. (i)  (ii)  (iii)
%% Conclusions. (i)  (ii)  (iii)

\end{abstract}

%%%%%%%%%%%%%
%%      INTRO	            %%
%%%%%%%%%%%%%

\section{Introduction}
Over the past two decades, a wide set of cosmological observations \citep{weinberg08} have allowed us to summarize our understanding of the
basic properties of the Universe in the concordance cosmological model, known as the $\Lambda$CDM model. Despite providing a good fit to the
observational data, the model presents some puzzles as most of the content of the Universe is in the form of \emph{dark} components, namely
dark matter and dark energy, whose nature is still mysterious.

In this framework, Cosmic Microwave Background (CMB) lensing science has emerged, in the last several years,  as a new promising cosmological probe \citep{smith:2007,das:2011,engelen:2012,planck_lens:2013,polarbar13,spt15,engelen:2015,PlanckCollaborationXV2015}. The large-scale structure (LSS) leaves an imprint on CMB anisotropies by gravitationally
deflecting CMB photons during their journey from the last-scattering surface to us \citep{bart01,lewis06}. The net effect is a remapping of the
CMB observables, dependent on the gravitational potential integrated along the line-of-sight (LOS). Thus the effect is sensitive to
both the geometry of the Universe and to the growth of the large-scale structure. Lensing also introduces non-Gaussian features in the CMB
anisotropy  pattern which  can be exploited to get information on the intervening mass distribution \citep{hu:2002,hirata:2003},  which in turn may give hints on the early stages of cosmic acceleration \citep{acquaviva2006, hu2006}.

On the other hand, since CMB lensing is an integrated quantity, it does not provide direct information on the \textit{evolution} of the large scale gravitational potential. However, the cross-correlation between CMB lensing maps and tracers of large-scale structure enables the  reconstruction of the dynamics and of the spatial distribution of the gravitational potential, providing simultaneous constraints on cosmological and astrophysical parameters \citep{pearson14}, such as the bias factor $b$ relating fluctuations in luminous and dark matter.

In the standard structure formation scenario galaxies reside in dark matter halos \citep{mo10} the most massive of which are the signposts
of larger scale structures that act as lenses for CMB photons. Bright sub-mm selected galaxies, that are thought to be the progenitors of present-day massive spheroidal galaxies, are excellent tracers of large-scale structure and thus optimally suited for cross-correlation studies.

Even more importantly, the sub-millimeter (sub-mm) flux density of certain sources remains approximately constant with increasing redshift for $z \gsim 1$ (strongly negative K-correction), so that sub-mm surveys have the power of piercing the distant Universe up to $z \gtrsim 3$--4 where the CMB lensing is most sensitive to matter fluctuations. In contrast, the available large--area optical/near-infrared galaxy surveys and radio source surveys reach redshifts only slightly above unity and therefore pick up only a minor fraction of the CMB lensing signal whose contribution peaks at $z > 1$ and is substantial up to much higher redshifts. Quasars allow us to extend investigations much further, but are rare and therefore provide a sparse sampling of the large-scale gravitational field.

Previous  cross-correlation studies involving CMB lensing and galaxy or quasar density maps have been reported by many authors
\citep{smith:2007,hirata:2008,bleem:2012,feng:2012,sherwin:2012,planck_lens:2013,geach:2013,giannantonio14,dipompeo15,bianchini15,allison15,omori15,desxc:2015,kuntz15,pullen15}.

As pointed out by \cite{song:2003}, the CMB lensing kernel is well-matched with the one of the unresolved dusty galaxies comprising the Cosmic Infrared Background (CIB) since both are tracers of the large-scale density fluctuations in the Universe. In particular, Planck measurements suggest that the correlation between  the CMB lensing map and
the CIB map at 545 GHz can be as high as $80\%$ \citep{planck_cib:2013}. Other statistically significant detections have been recently reported by \cite{holder:2013,hanson:2013,polarberar_herschel:2014,engelen:2015}. Even though there are connections between these studies and the one presented here, one needs to bear in mind that, differently from galaxy catalogs, the CIB is an integrated quantity and as such it prevents a detailed investigation of the temporal evolution of the signal. Moreover, the interpretation of the measured cross-correlation is actually more challenging since the precise redshift distribution of the sources contributing to the sub-mm background is still debated.

In this paper we revisit the angular cross-power spectrum $C_{\ell}^{\kappa g}$ between the CMB convergence derived from \emph{Planck} data and the spatial distribution of high-$z$ sub-mm galaxies detected by the \emph{Herschel}\footnote{\textit{Herschel} is an ESA space observatory with science instruments provided by European-led Principal Investigator consortia and with an important participation from NASA.} Astrophysical Terahertz Large Area Survey \citep[H-ATLAS;][]{eales10}. The present analysis improves over that presented in our  previous paper \citep[][hereafter B15]{bianchini15} in several aspects: (i) we adopt the new \textit{Planck} CMB lensing map \citep[][see Sect.~\ref{subsec:planck}]{PlanckCollaborationXV2015}; (ii) we treat more carefully the uncertainty in the photometric redshift estimates of the H-ATLAS galaxy sample (see Sect.~\ref{subsec:herschel}); (iii) we move toward a tomographic study of the cross-correlation signal (see Sect.~\ref{subsec:tomography}).

The outline of this paper is as follows: in Section~\ref{sec:theory} we briefly review the theoretical background, in Section~\ref{sec:data}
we introduce the datasets, while the analysis method is presented in Section~\ref{sec:analysis}. In Section~\ref{sec:results} we report and
analyze the derived constraints on the galaxy bias parameter, discussing potential systematic effects that can affect the cross-correlation. Finally in Section \ref{sec:conclusions} we summarize our results.

Throughout this paper we adopt the fiducial flat $\Lambda$CDM cosmology with best-fit \emph{Planck} + WP + highL + lensing cosmological parameters as provided by \citet{planck_parameters:2013}. Here, WP refers to WMAP polarization data at low multipoles, highL to the inclusion of high-resolution CMB data of the Atacama Cosmology Telescope (ACT) and South Pole Telescope (SPT) experiments, and lensing to the inclusion of \emph{Planck} CMB lensing data in the parameter likelihood.

%%%%%%%%%%%%%
%         THEORY               %
%%%%%%%%%%%%%
\section{Theory}
\label{sec:theory}
Both the CMB convergence field $\kappa$ and the galaxy density fluctuation field $g$ along the LOS can be written as a weighted integral  over redshift of the dark
matter density contrast $\delta$:

\begin{equation}
X(\nver) = \int_0^{z_*} dz\, W^X(z)\delta(\chi(z)\nver,z),
\end{equation}
where $X=\{\kappa,g\}$ and $W^X(z)$ is the kernel related to a given field.
The kernel $W^{\kappa}$, describing the lensing efficiency of the matter distribution, writes
\begin{equation}
W^{\kappa}(z) = \frac{3\Omega_m}{2c}\frac{H_0^2}{H(z)}(1+z)\chi(z)\frac{\chi_*-\chi(z)}{\chi_*}.
\end{equation}
Here, $\chi(z)$ is the comoving distance to redshift $z$, $\chi_*$ is the comoving distance to the last scattering surface at $z_*\simeq
1090$, $H(z)$ is the Hubble factor at redshift $z$, $c$ is the speed of light, $\Omega_m$ and $H_0$ are the present-day values of matter
density and Hubble parameter, respectively.

Assuming that the luminous matter traces the peaks of the dark matter distribution, the galaxy kernel is given by the sum of two terms:
\begin{equation}
W^{g}(z) = b(z)\frac{dN}{dz} + \mu(z).
\label{eqn:wg}
\end{equation}
The first term is related to the physical clustering of sources and is the product of the bias factor\footnote{Throughout the analysis we assume a linear, local, deterministic, redshift- and scale-independent bias factor unless otherwise stated.} $b$ with the \emph{unit-normalized} redshift distribution of sources, $dN/dz$. The second term describes the effect of the lensing magnification bias \citep{turner84,villumsen95,xia09}; it writes:
\begin{equation}
\label{eqn:wmu}
\begin{split}
\mu(z) &= \frac{3\Omega_{\rm m}}{2c}\frac{H_0^2}{H(z)}(1+z)\chi(z) \\
&\times \int_z^{z_*}dz'\,\Bigl(1-\frac{\chi(z)}{\chi(z')}\Bigr)(\alpha(z')-1)\frac{dN}{dz'}.
\end{split}
\end{equation}
This term is independent of the bias parameter and, in the weak lensing limit, depends on the slope of the galaxy number counts $\alpha$ ($N(>S)\propto S^{-\alpha}$) at the flux density limit of the survey. \cite{gonzalez-nuevo:2014} have shown that the magnification bias by weak lensing is substantial for high-$z$ H-ATLAS sources selected with the same criteria as the present sample. In this analysis the value is estimated from the data, at flux densities immediately above the flux density limit;  we find $\alpha\simeq 3$ and fix it to this value.

The theoretical CMB convergence-galaxy angular cross-power spectrum and the galaxy auto-power spectrum can be evaluated in the Limber
approximation \citep{limber} as a weighted integral of the matter power spectrum $P(k,z)$:
\begin{equation}\label{eq:cross}
\begin{split}
C_{\ell}^{\kappa g} &=   \int_0^{z_*} \frac{dz}{c} \frac{H(z)}{\chi^2(z)} W^{\kappa}(z)W^{g}(z)P(k=\ell/\chi(z),z); \\
C_{\ell}^{gg} &=   \int_0^{z_*} \frac{dz}{c} \frac{H(z)}{\chi^2(z)} [W^{g}(z)]^2P(k=\ell/\chi(z),z).
\end{split}
\end{equation}
We compute the non-linear $P(k,z)$ using the \lstinline!CAMB!\footnote{\url{http://cosmologist.info/camb/}} code with the \lstinline!Halofit!
prescription \citep{camb,halofit}. \\ The expected signal-to-noise (S/N) of the detection for the CMB convergence-density correlation can be estimated assuming that both fields behave as Gaussian random fields, so that the variance of $C_{\ell}^{\kappa g}$ is

\begin{equation}
\label{eqn:delta_kg}
\bigl(\Delta C_{\ell}^{\kappa g}\bigr)^2 = \frac{1}{(2\ell+1)f_{\rm sky}} \bigl[(C_{\ell}^{\kappa g})^2 + (C_{\ell}^{\kappa\kappa}+N_{\ell}^{\kappa\kappa})(C_{\ell}^{gg}+N_{\ell}^{gg})\bigr],
\end{equation}

where $f_{\rm sky}$ is the sky fraction observed by both the galaxy and the lensing surveys, $N_{\ell}^{\kappa\kappa}$ is the CMB lensing reconstruction noise level, $N_{\ell}^{gg}=1/\bar{n}$ is the shot noise associated with the galaxy field, and $\bar{n}$ is the mean number of sources per steradian. For the H-ATLAS - \emph{Planck} CMB lensing cross-correlation the sky coverage is approximately 600 deg$^2$ ($f_{\rm sky}\simeq 0.01$) and we assume a constant bias $b=3$ and a slope of the galaxy number counts $\alpha=3$: restricting the analysis between $\ell_{min}=100$ (as lower multipoles are poorly reconstructed) and $\ell_{max}=800$, we expect $S/N \simeq 7.5$.

%%%%%%%%%%%%%
%  DATA & ANALYSIS   %
%%%%%%%%%%%%%

\section{Data}
\label{sec:data}
\subsection{\emph{Planck} data}
\label{subsec:planck}
We make use of the publicly released 2015 \emph{Planck}\footnote{Based on observations obtained with the \emph{Planck} satellite
(\url{http://www.esa.int/Planck}), an ESA science mission with instruments and contributions directly funded by ESA Member States, NASA, and
Canada.} CMB lensing map \citep{PlanckCollaborationXV2015} that has been extracted by applying a quadratic estimator \citep{okamoto:2003} to
foreground-cleaned temperature and polarization maps. These maps have been synthesized from the raw 2015 \emph{Planck} full mission frequency maps using the \lstinline!SMICA! code \citep{PlanckCollaborationIX2015}. In particular, the released map is based on a minimum-variance (MV) combination of all temperature and polarization estimators, and is provided as a mean-field bias subtracted convergence $\kappa$ map.

For a comparison, we also use the earlier CMB lensing data provided within the \emph{Planck} 2013 release \citep{planck_lens:2013}. Differently from the 2015 case, the previous lensing map is
based on a MV combination of only the 2013 \emph{Planck} 143 and 217 GHz  foreground-cleaned temperature anisotropy maps. The lensing mask associated to the 2015 release covers a slightly larger portion of the sky with respect to the 2013 release: $f_{\rm sky}^{2015}/f_{\rm sky}^{2013}\simeq0.98$.

The exploitation of the full-mission temperature and the inclusion of polarization data have the effect of augmenting the \emph{Planck} lensing reconstruction sensitivity. Both maps are in the \lstinline!HEALPix!\footnote{\url{http://healpix.jpl.nasa.gov}} \citep{healpix} format with a resolution parameter $N_{\rm side} =
2048$. We downgraded them to a resolution of $N_{\rm side}=512$ (corresponding to an angular resolution of $\sim 7'.2$).

\subsection{\textit{Herschel} data} \label{subsec:herschel}
The H-ATLAS \citep{eales10} is the largest extragalactic key-project carried out in open time with the \textit{Herschel} Space Observatory \citep{pilbratt10}. It was allocated 600 hours of observing time and covers about $600\,\hbox{deg}^2$ of sky in five photometric bands (100, 160, 250, 350 and $500\,\mu$m) with the Photodetector Array Camera and Spectrometer \citep[PACS; ][]{poglitsch10} and the Spectral and Photometric Imaging Receiver \citep[SPIRE;][]{griffin10}. The H-ATLAS map-making is described by \citet{Pascale2011} for SPIRE and by \citet{Ibar2010} for PACS. The procedures for source extraction and catalogue generation can be found in \citet{Rigby2011} and \citet{Valiante2015}.

Our sample of sub-mm galaxies is extracted from the same internal release of the full H-ATLAS catalogue as in B15. The survey area is divided into five fields: the north galactic pole (NGP), the south galactic pole (SGP) and the three GAMA fields (G09, G12, G15). The H-ATLAS galaxies have a broad redshift distribution extending from $z=0$ to $z\simeq 5$ \citep{pearson13}. The $z \lesssim 1$ population is mostly made of ``normal'' late-type and star-burst galaxies with low to moderate star formation rates  \citep[SFRs;][]{dunne11,guo11} while the high-$z$ galaxies are forming stars at high rates ($\hbox{SFR}\gtrsim \hbox{few hundred}\,M_{\odot}\,\hbox{yr}^{-1}$) and are much more strongly clustered \citep{Maddox2010,Xia2012}, implying that they are tracers of large-scale overdensities. Their properties are consistent with them being the progenitors of local massive elliptical galaxies \citep{lapi11}.

We have selected a sub-sample of H-ATLAS galaxies complying with the following criteria: (i) flux density at $250\,\mu$m, $S_{250}> 35$ mJy; (ii) $\ge 3\,\sigma$ detection at $350\,\mu$m; and (iii) photometric redshift greater than a given value, $z_{\rm ph, min}$, as discussed below. For our baseline analysis we set $z_{\rm ph, min} = 1.5$. The sample comprises 94,825 sources. It was subdivided into two redshift bins ($1.5 \le z_{\rm ph} < 2.1$ and $z_{\rm ph} \ge 2.1$) containing a similar number of sources (53,071 and 40,945, respectively).

\begin{figure*} 
\plotone{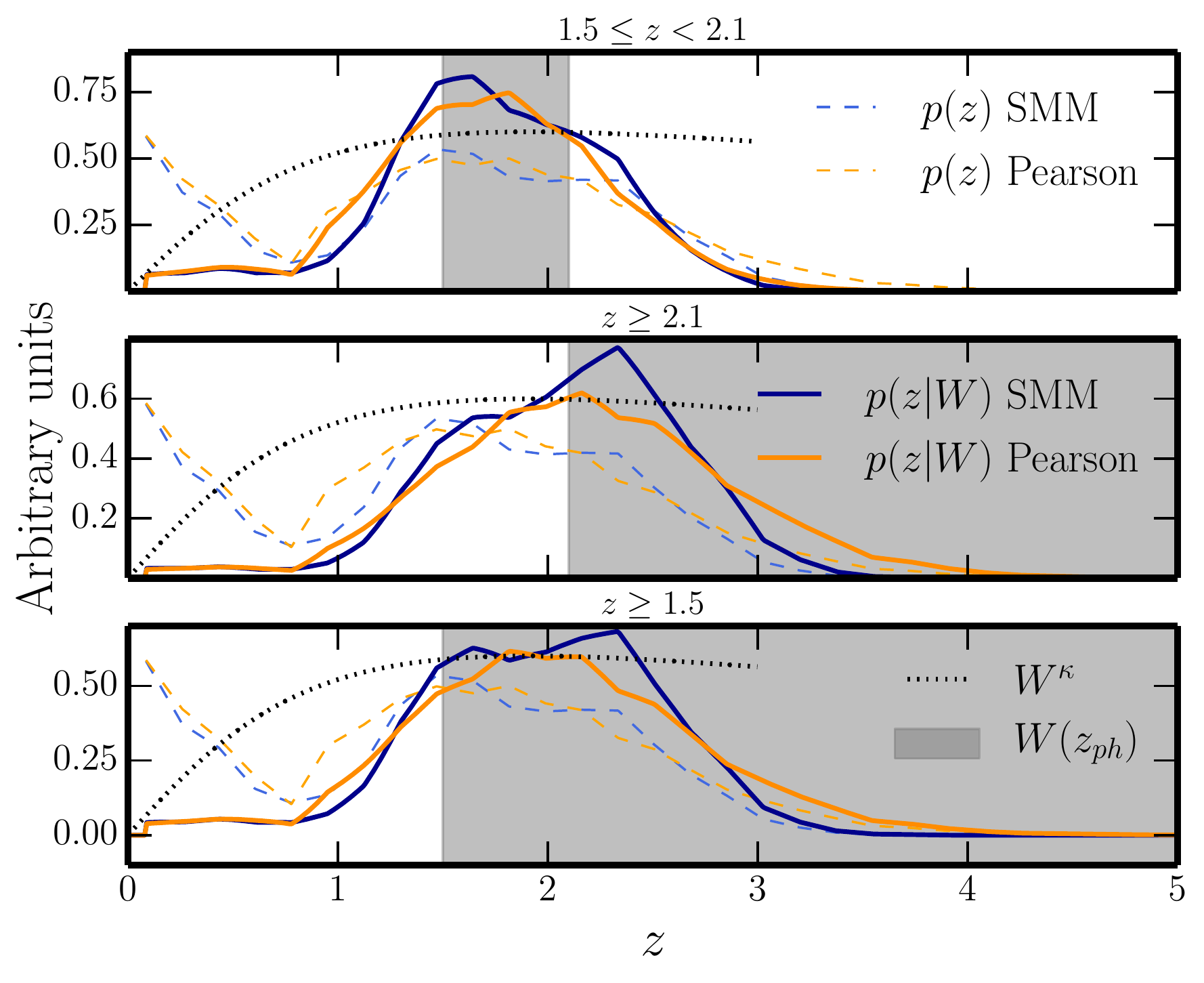}
\caption{Redshift distributions of the galaxy samples selected from the full H-ATLAS survey catalogue. The fiducial  \emph{true} redshift distribution $p(z)$ for the full sample is shown, in each panel, by the dashed lines, while the solid lines show the redshift distributions $p(z|W)$ obtained implementing the top-hat window functions $W(z_{\rm ph})$ represented, in each panel, by the shaded area. The blue and the orange lines refer to redshift distributions based on the SED of SMM~J2135-0102 and on the \citet{pearson13} best fitting template, respectively (see Sect.~\ref{sec:SED}). These redshift distributions were used for the evaluation of the theoretical $C_{\ell}$'s. The dotted black line in each panel shows the arbitrarily normalized CMB lensing kernel $W^{\kappa}(z)$.
  \label{fig:dndz}}
\end{figure*}

The estimate of the \emph{unit-normalized} redshift distributions $dN/dz$ to be plugged into eq.~(\ref{eqn:wg}), hence into eq.~(\ref{eq:cross}), is a very delicate process because of the very limited spectroscopic information. Also, the fraction of H-ATLAS sources having accurate photometric redshift determinations based on multi-wavelength optical/near-infrared photometry is rapidly decreasing with increasing redshift above $z\simeq 0.4$ \citep{smith11,Bourne2015}. However, if a typical rest-frame spectral energy distribution (SED) of H-ATLAS galaxies can be identified, it can be used to estimate the redshifts directly from \textit{Herschel} photometric data.

\citet{lapi11} and \citet{gnuevo12} showed that the SED of SMM~J2135-0102, `The Cosmic Eyelash' at $z=2.3$ \citep{Ivison2010,Swinbank2010} is a good template for $z\gtrsim 1$. Comparing the photometric redshift obtained with this SED with spectroscopic measurements for the 36 H-ATLAS galaxies at $z\gtrsim 1$ for which spectroscopic redshifts were available \citet{gnuevo12} found a median value of $\Delta z/(1+z)\equiv (z_{\rm phot}-z_{\rm spec})/(1+z_{\rm spec})=-0.002$ with a dispersion $\sigma_{\Delta z/(1+z)}= 0.115$. At lower redshifts this template performs much worse. As argued by \citet{lapi11} this is because the far-IR/sub-mm SEDs of H-ATLAS galaxies at $z>1$ are dominated by the warm dust component while the cold dust component becomes increasingly important with decreasing $z$, amplifying the redshift--dust temperature degeneracy. That's why we restrict our analysis to $z_{\rm ph}\ge1.5$.

\citet{pearson13} generated an average template for $z>0.4$  H-ATLAS sources using a subset of 53 H-ATLAS sources with measured redshifts in the range $0.4 < z < 4.2$. They found that the redshifts estimated with this template have an average offset from spectroscopic redshift of $\Delta z/(1+z)=0.018$ with a dispersion  $\sigma_{\Delta z/(1+z)}= 0.26$.

\begin{center} 
\begin{deluxetable*}{cccccccc}
\tabletypesize{}
%\rotate
\tablecaption{Statistics of H-ATLAS fields\label{herschel_patches}}
\tablewidth{0pt}
\tablehead{
\colhead{} & \multicolumn{3}{c}{$N_{\rm obj}$} & \colhead{} & \multicolumn{3}{c}{$\bar{n}$ [gal ster$^{-1}$]}\\
\cline{2-4} \cline{6-8} \\
\colhead{Patch} & \colhead{$z_{\rm ph} \ge 1.5$} & \colhead{${1.5 \le z_{\rm ph} < 2.1}$} & \colhead{$z_{\rm ph} \ge 2.1$}
& \colhead{} & \colhead{$z_{\rm ph} \ge 1.5$} & \colhead{${1.5 \le z_{\rm ph} < 2.1}$} & \colhead{$z_{\rm ph} \ge 2.1$}
}
\startdata
ALL   &   94825 &   53071   &    40945   &  & $5.76 \cdot 10^{5}$ & $3.22 \cdot 10^{5}$ & $2.49 \cdot 10^{5}$\\
NGP  &   26303 &   15033   &    11039  & & $5.63 \cdot 10^{5}$ & $ 3.22 \cdot 10^{5}$ & $2.36 \cdot 10^{5}$\\
SGP  &   43518 &   24722   &    18422  & & $5.95 \cdot 10^{5}$ & $3.38 \cdot 10^{5}$ & $2.52 \cdot 10^{5}$  \\
G09   &   8578   &   4590   &    3922    & & $5.72 \cdot 10^{5}$ & $3.02 \cdot 10^{5}$ & $2.61 \cdot 10^{5}$\\
G12   &   8577   &   4611   &    3881    & & $5.34 \cdot 10^{5}$ & $2.87 \cdot 10^{5}$ & $2.41 \cdot 10^{5}$\\
G15   &   7849   &   4115   &    3681    & & $5.66 \cdot 10^{5}$ & $2.97 \cdot 10^{5}$ & $2.65 \cdot 10^{5}$\\
\enddata
\tablenotetext{a}{ALL is  the combination of all the fields together.}
\end{deluxetable*}
\end{center}

In the following we will use the SED of SMM~J2135-0102 as our baseline template; the effect of using the template by \citet{pearson13} is presented Sect.~\ref{sec:SED}. To allow for the effect on $dN/dz$ of random errors in photometric redshifts we estimated, following \citet{budavari03}, the redshift distribution, $p(z|W)$, of galaxies selected by our window function $W(z_{\rm ph})$,  as
\begin{equation}
p(z|W) = p(z) \int dz_{\rm ph} W(z_{\rm ph})p(z_{\rm ph}|z),
\end{equation}
where $p(z)$ is the fiducial true redshift distribution, $W=1$ for $z_{\rm ph}$ in the selected interval and $W=0$ otherwise, and $p(z_{\rm ph}|z)$ is probability that a galaxy with a true redshift $z$ has a photometric redshift $z_{\rm ph}$. The error function $p(z_{\rm ph}|z)$ is parameterized as a Gaussian distribution with zero mean and dispersion $(1+z)\,\sigma_{\Delta z/(1+z)}$. For the dispersion we adopt the conservative value $\sigma_{\Delta z/(1+z)}=0.26$.

A partial estimate of the effect of the dust temperature--redshift degeneracy in contaminating our $z> z_{\rm ph, min}$ sample derived from \textit{Herschel} colors by cold low-$z$ galaxies is possible thanks to the work by \citet{Bourne2015} who used a likelihood-ratio technique to identify SDSS counterparts at $r < 22.4$ for H-ATLAS sources in the GAMA fields and collected spectroscopic and photometric redshifts from GAMA and other public catalogues. A cross-match with their data set showed that about 7\% of sources in GAMA fields with estimated redshifts larger than 1.5 based on \textit{Herschel} colors have a reliable ($R \ge 0.8$) optical/near-IR counterpart with photometric redshift $<1$. The fiducial redshift distribution for the GAMA fields was corrected by moving these objects and the corrected, unit normalized, redshift distribution was adopted for the full sample. The result is shown in Fig.~\ref{fig:dndz} for $z\ge 1.5$ and for the sub-sets at $1.5\le z_{\rm ph} \le 2.1$ and $z_{\rm ph} > 2.1$. As mentioned above, photometric redshifts based on \textit{Herschel} colors become increasingly inaccurate below $z\sim 1$. Thus the low-$z$ portions of the $p(z)$'s in Fig.~\ref{fig:dndz} are unreliable.

For each H-ATLAS field we created an overdensity map in \lstinline!HEALPix! format with a resolution parameter $N_{\rm side}=512$. The overdensity is defined as $g(\nver)= n(\nver)/\bar{n}-1$, where $n(\nver)$ is the number of objects in a given pixel, and $\bar{n}$ is the mean number of objects per pixel. As a last step, we combined the \emph{Planck} lensing mask with the H-ATLAS one. The total sky fraction retained for the analysis is $f_{\rm sky}=0.013$. The specifics of each patch are summarized in Table~\ref{herschel_patches}.

\section{Analysis method}
\label{sec:analysis}
\subsection{Estimation of the power spectra}
\label{subsec:methodology}
We measured the cross-correlation between the \emph{Planck} CMB lensing convergence and the H-ATLAS galaxy overdensity maps in the harmonic
domain. Unbiased (but slightly sub-optimal) bandpower estimates are obtained using a pseudo-$C_{\ell}$ method based on the \lstinline!MASTER!
algorithm \citep{master}.  The estimator of the true band powers $\hat{C}^{\kappa g}_{L}$ writes
\begin{equation}
\label{eqn:master_kg}
\hat{C}^{\kappa g}_{L} = \sum_{L' \ell}K^{-1}_{LL'}P_{L'\ell}\tilde{C}^{\kappa g}_{\ell},
\end{equation}
where $\hat{C}$ denotes the observed power spectrum, $\tilde{C}$ denotes the pseudo-power spectrum, and $L$ is the bandpower index
(hereafter $C^{XY}_{L}$ denotes the binned power spectrum and  $L$ identifies the bin). The binned coupling matrix can be written as
\begin{equation}
K_{LL'} = \sum_{\ell\ell'} P_{L\ell}M_{\ell\ell'}B^2_{\ell'}Q_{\ell' L'}.
\end{equation}
Here $P_{L\ell}$ is the binning operator; $Q_{\ell L}$ and $B^2_{\ell'}$ are, respectively, the reciprocal of the binning operator and  the
pixel window function that takes into account the finite pixel size. By doing so, we take into account the mode-coupling induced by the
complex geometry of the survey mask and correct for the pixel window function.  The signal is estimated in seven linearly spaced bins of
width $\Delta\ell = 100$ spanning the multipole range from 100 to 800. A thorough description of the pipeline implementation and validation
can be found in B15, where a comparison between different error estimation methods is also given.

The auto-correlation signal is extracted with the same procedure. However, in the case of the galaxy auto-power spectrum, we have to subtract from the estimated bandpowers the shot-noise term $N^{gg}_{\ell}=1/\bar{n}$.

\begin{figure} 
\plotone{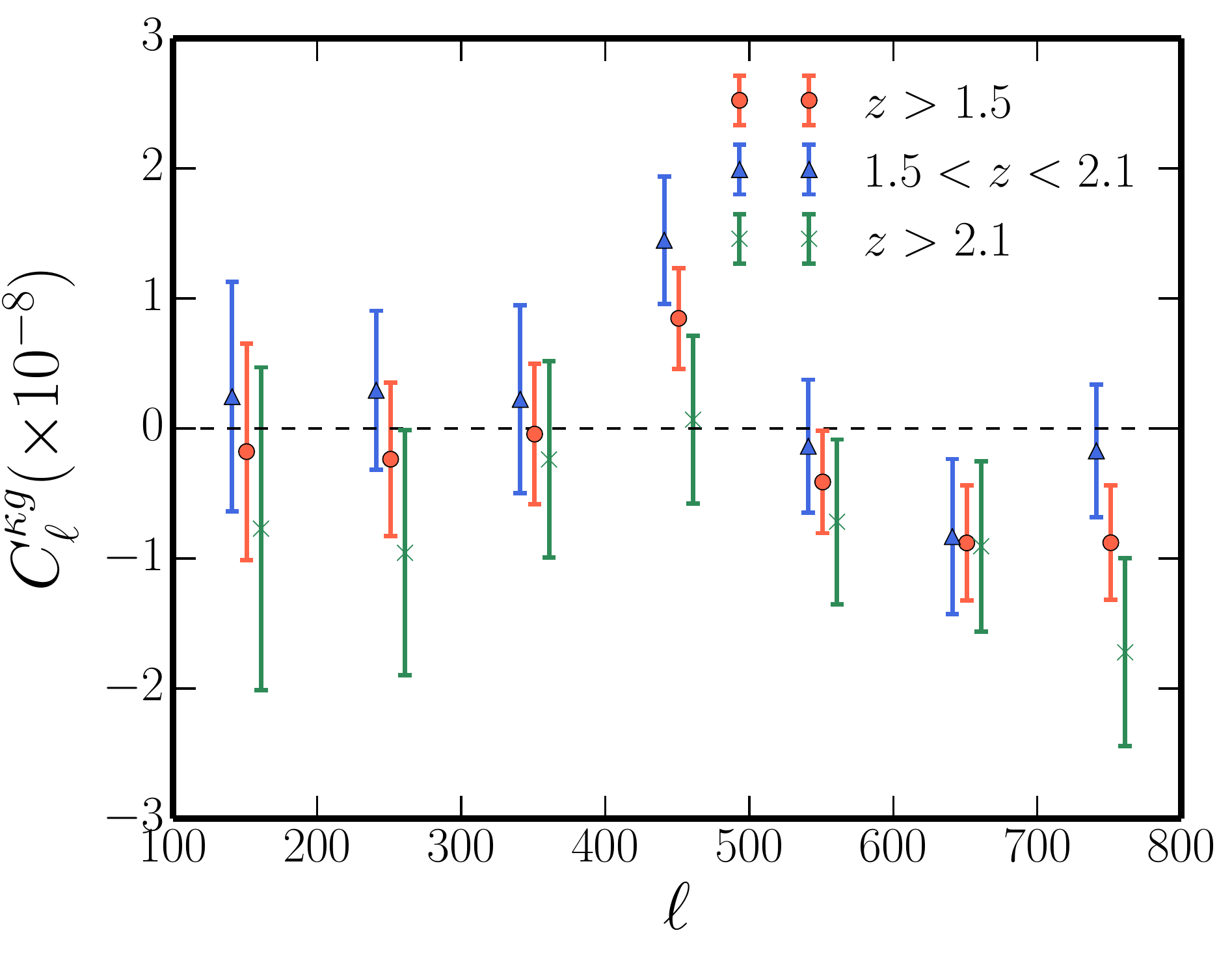}
\caption{Null test results. Mean cross-spectrum $C^{\kappa g}_{\ell}$ between $N_{\rm sim}=100$ simulated \emph{Planck} CMB lensing maps and
the H-ATLAS galaxy density maps for the three redshift bins considered: $z \ge 1.5$ (blue circles), $1.5 \le z < 2.1$ (green crosses), and
$z \ge 2.1$ (red triangles). Band-powers are displaced by $\Delta\ell = \pm 10$ with respect to the bin centers for visual clarity. The error bars are calculated as the square root of the covariance matrix diagonal derived from the same set of simulations and divided by $\sqrt{N_{\rm sim}}$.  \label{fig:null}}
\end{figure}

In order to estimate the full covariance matrix and the error bars we make use of the publicly available set of 100 realistic CMB convergence
simulations\footnote{\url{http://irsa.ipac.caltech.edu/data/Planck/release_2/all-sky-maps/maps/component-maps/lensing}}, that accurately reflect the \emph{Planck} 2015 noise properties, and cross-correlate them with the H-ATLAS galaxy density contrast maps. Because there is no correlated cosmological signal between CMB lensing simulations and real galaxy datasets, we also use them to check that our pipeline does not introduce any spurious signal. The mean cross-spectrum between the \emph{Planck} simulations and the H-ATLAS maps is shown in Fig.~\ref{fig:null} which shows that it is consistent with zero in all redshift bins. For the baseline photo-$z$ bin we obtain $\chi^2 = 9.5$ for $\nu=7$ degrees-of-freedom (d.o.f.), corresponding to a probability of random deviates with the same covariances to exceed this chi-squared (p-value) of 0.22. In the other two redshift bins we find $\chi^2=12.6$ for the low-$z$ one and $\chi^2=6.1$ for the high-$z$ one, corresponding to a p-value $p=0.08$ and $p=0.53$ respectively.

\subsection{Estimation of the cross-correlation amplitude and of the galaxy bias}
\label{subsec:mcmc}

Following B15 we introduce a phenomenologically-motivated amplitude parameter $A$ that globally scales the observed cross-power spectrum
with respect to the theoretical one as $\hat{C}^{\kappa g}_{L}=AC^{\kappa g}_{L}(b)$. We analyze the constraints on the parameters $A$ and $b$ combining the information from the cross-spectrum and the galaxy auto-spectrum. For the joint analysis we exploit Gaussian likelihood functions that take into account correlations between the cross- and the auto-power spectra in the covariance matrix. The extracted cross- and auto-band-powers are organized into a single data vector as
\begin{equation}
\mathbf{\hat{C}}_{L} = (\mathbf{\hat{C}}^{\kappa g}_{L}, \mathbf{\hat{C}}^{gg}_{L}),
\end{equation}
which has $N_L=14$ elements. The total covariance matrix is then written as the composition of four $7\times7$ submatrices:
\begin{equation}
\textbf{Cov}_{LL'} =
\begin{bmatrix}
\text{Cov}^{\kappa g}_{LL'} & (\text{Cov}^{\kappa g-gg}_{LL'})^\intercal \\
 \text{Cov}^{\kappa g-gg}_{LL'} & \text{Cov}^{gg}_{LL'}  \\
\end{bmatrix}.
\end{equation}
The covariance matrices are approximately given by:
\begin{equation}
\label{eqn:covs}
\begin{split}
{\text{Cov}}^{gg}_{LL'} &= \frac{2}{(2L+1)\Delta L f_{\rm sky}}\Bigl[C_{L}^{gg}(\boldsymbol{\theta}) + N_{L}^{gg}\Bigr]^2\delta_{LL'}
; \\
{\text{Cov}}^{\kappa g}_{LL'} &= \frac{1}{(2L+1)\Delta L f_{\rm sky}} \\
&\!\!\!\!\!\!\!\!\!\!\!\!\!\!\times\Bigl[(C_{L}^{\kappa g}(\boldsymbol{\theta}))^2 + (C_{L}^{\kappa\kappa} + N_{L}^{\kappa\kappa})(C_{L}^{gg}(\boldsymbol{\theta}) +
N_{L}^{gg})\Bigr] \delta_{LL'} ; \\
{\text{Cov}}^{\kappa g-gg}_{LL'} &\!\!\!=\! \frac{2}{(2L\!+\!1)\Delta L f_{\rm sky}} \Bigl [(C^{gg}_{L}(\boldsymbol{\theta})\!+\!N^{gg}_{L})C^{\kappa
g}_{L}(\boldsymbol{\theta}) \Bigr] \delta_{LL'},
\end{split}
\end{equation}
where the $\Delta L$ is the bin size, $\boldsymbol{\theta}$ is the parameters vector, and
$\delta_{LL'}$ is the Kronecker delta such that they are diagonal.
Then, the likelihood function can be written as
\begin{equation}
\begin{split}
\mathcal{L}&(\mathbf{\hat{C}}_{L}|\boldsymbol{\theta}) = (2\pi)^{-N_L/2} [\text{det}\, \textbf{Cov}_{LL'}]^{-1/2} \\
\!\!\!\!\!\!&\!\!\!\!\!\!\times \exp{ \Bigl\{ -\frac{1}{2} \bigl[ \mathbf{\hat{C}}_{L} - \mathbf{C}_{L}(\boldsymbol{\theta})  \bigr]
\bigl(\textbf{Cov}_{LL'}\bigr)^{-1} \bigl[ \mathbf{\hat{C}}_{L'} - \mathbf{C}_{L'}(\boldsymbol{\theta})  \bigr]  \Bigr\}}.
\end{split}
\end{equation}
The parameter space is explored using \lstinline!emcee!\footnote{\url{http://dan.iel.fm/emcee}}, an affine invariant Markov Chain Monte
Carlo (MCMC) sampler \citep{emcee}, assuming flat priors over the ranges $\{b,A, \mathcal{A}_{\rm bias}\} = \{[0,10], [-1, 10], [0, 10] \}$\footnote{Note that we constrain separately $(b,A)$ and $(A, \mathcal{A}_{\rm bias})$.} ($\mathcal{A}_{\rm bias}$ will be defined in Sect.~\ref{sec:bz}). This analysis scheme is applied independently to each redshift bin.

The covariance matrices built with the 100 \textit{Planck} lensing simulations were used to compute the error bars for the cross-power spectra (the ones shown in \cref{fig:null,fig:kg_data,fig:kg_data_tomo,fig:kg_data_tomo_Pearson}), to address bin-to-bin correlations and to evaluate the chi-square for the null-hypothesis rejection. On the other hand, we used the diagonal analytical approximation of Eq.~(\ref{eqn:covs}) to evaluate the bias-dependent covariance matrices used to sample the posterior distribution and for error bars on the galaxy power spectra estimation (error bars shown in \cref{fig:gg_data,fig:gg_data_tomo,fig:gg_cross,fig:gg_data_tomo_Pearson}). As in B15, we decided to rely on an analytical approximation of the covariance matrices that depend on the estimated parameters, i.e. the linear galaxy bias. This simple approximation is able to capture the covariance matrices features as shown in Sec.~\ref{subsec:tomography} where we compare results obtained with (i) the diagonal approximation given by Eq.~\ref{eqn:covs}; (ii) the non-diagonal (bias-dependent) analytical prescription derived and exploited in B15 that accounts for the mask induced mode-coupling; (iii) and the full covariance matrix evaluated from the set of 100 Planck CMB lensing simulations as described above.

%%%%%%%%%%%%%
%%        RESULTS      %%
%%%%%%%%%%%%%

\section{Results and discussion}
\label{sec:results}
\subsection{Comparison between the 2013 and the 2015 \emph{Planck} results}
\begin{figure} 
\plotone{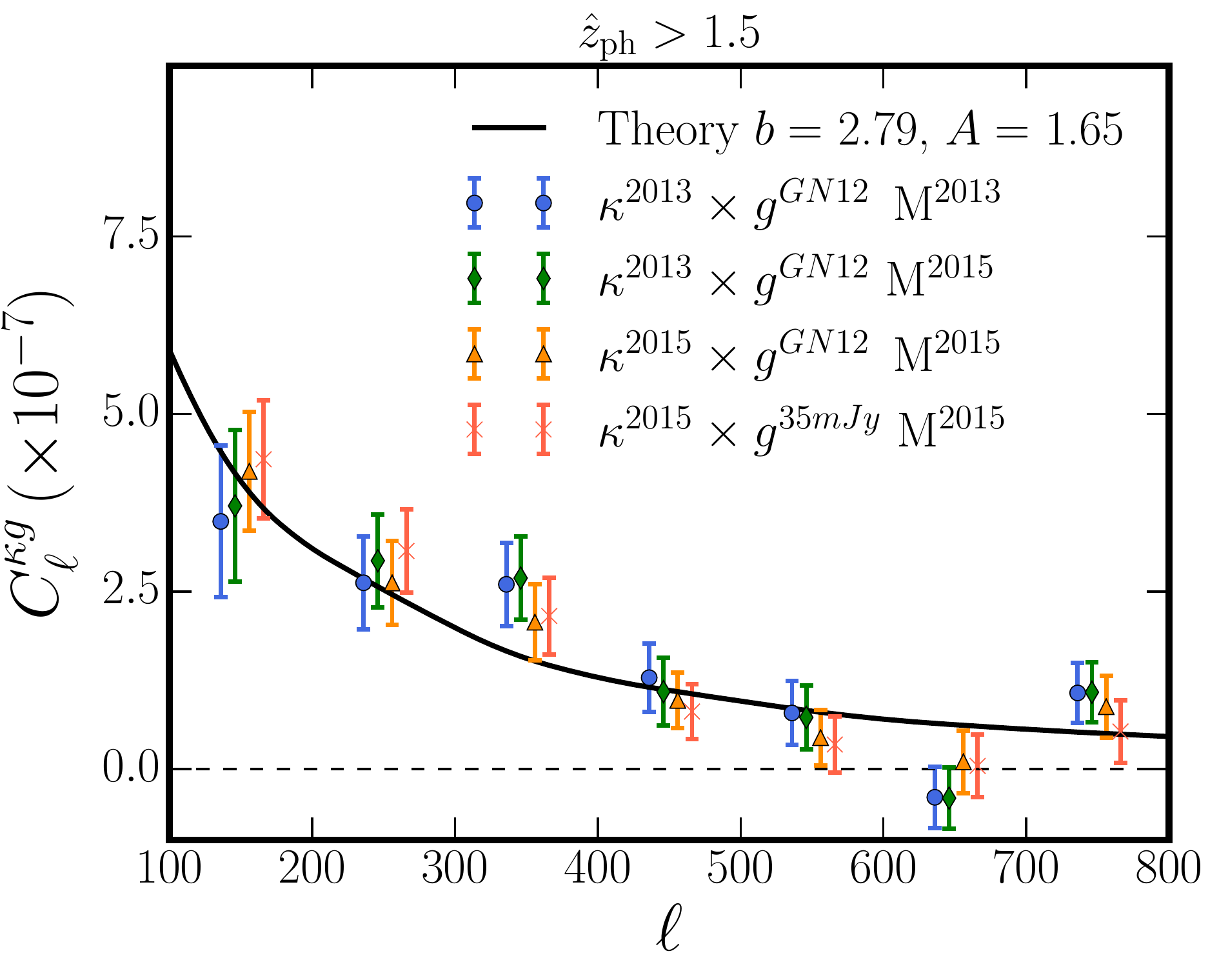}
\caption{Comparison of the cross-spectra between the \emph{Planck} CMB lensing maps and the $z_{\rm ph} \ge 1.5$ H-ATLAS galaxy density maps obtained using the 2013 and the 2015 \emph{Planck} results. The results labelled ``$[k^{2015}\times g^{35\rm mJy}]M^{2015}$'' refer to the analysis done with the present sample of $z_{\rm ph}\ge1.5$ galaxies. The GN12 superscript refers to the sample of H-ATLAS galaxies used by B15, based on slightly more restrictive selection criteria; results for this sample are shown for both the 2013 or the 2015 \emph{Planck} masks ($M^{2013}$ and $M^{2015}$) and convergence maps ($\kappa^{2013}$ and $\kappa^{2015}$). The solid black line shows the theoretical cross-spectrum for the best-fit values of the bias factor and of the cross-correlation amplitude, $A$, found for the $[\kappa^{2015}\times g^{35\rm mJy}]M^{2015}$ adopting the redshift distribution  estimated by B15. The estimated bandpowers are plotted with an offset along the $x$-axis for a better visualization. The error-bars were computed using the full covariance matrix obtained via Monte Carlo simulations as $\Delta C^{\kappa g}_L = \sqrt{\text{diag}(\text{Cov}^{\kappa g})}$.
  \label{fig:kg_data}}
\end{figure}

Figure~\ref{fig:kg_data} compares the cross-spectra $C_{\ell}^{\kappa g}$ between the $z_{\rm ph} \ge 1.5$ H-ATLAS galaxy sample and the 2013/2015 \emph{Planck} CMB lensing maps. For a fuller comparison, the figure also shows the results obtained using the galaxy catalogue built by B15 adding to the requirements (i-iii) mentioned in Sect.~\ref{subsec:herschel} the color criteria introduced by \cite{gnuevo12} (hereafter GN12): $S_{350\,\mu\rm m}/S_{250\,\mu\rm m}>0.6$ and $S_{500\,\mu\rm m}/S_{350\,\mu\rm m}>0.4$. The error bars were derived by cross-correlating the 100 simulated \emph{Planck} lensing realizations with the sub-mm galaxy map and measuring the variance in $C_{\ell}^{\kappa g}$.

\begin{figure} 
\plotone{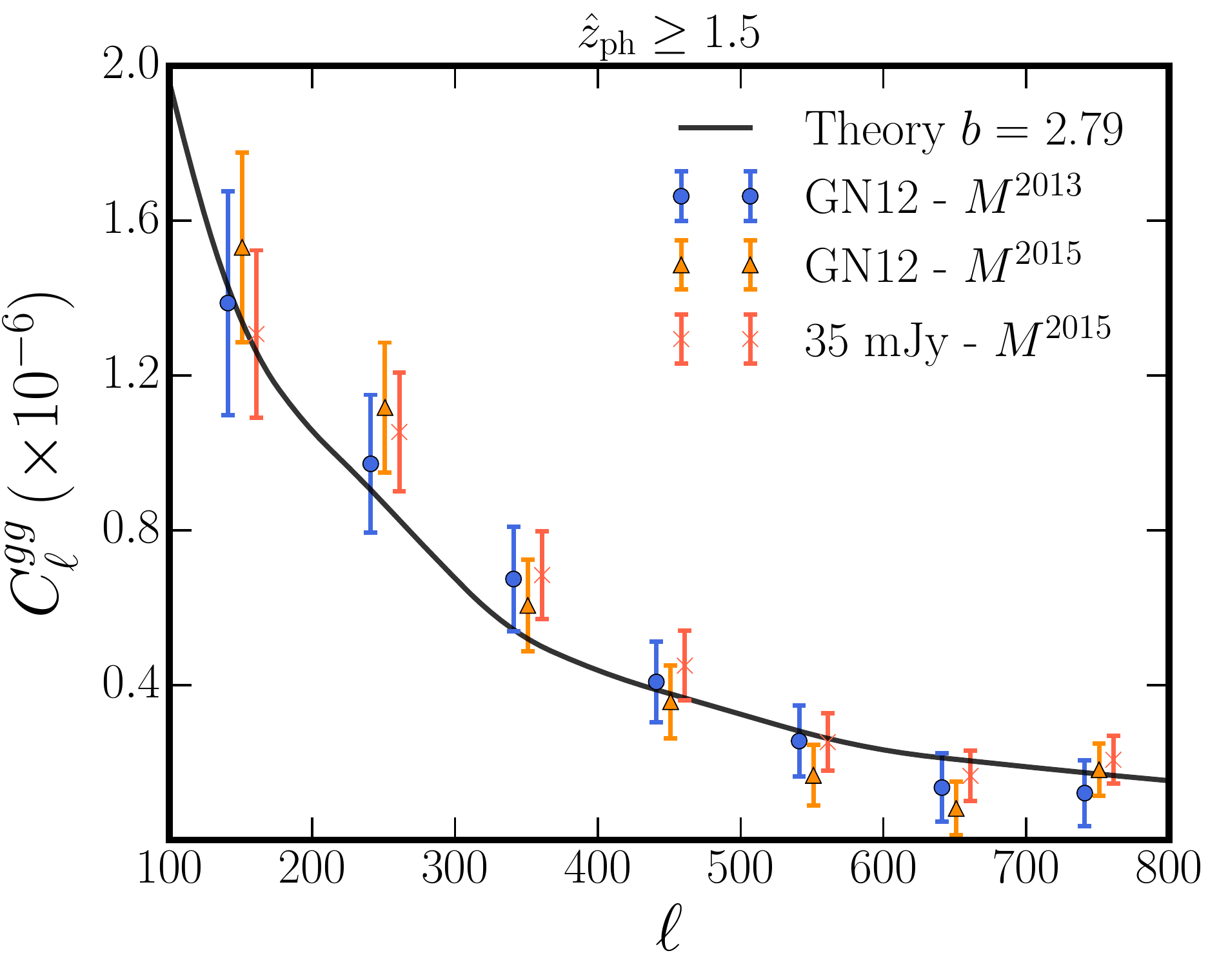}
\caption{Comparison of the auto-power spectra of H-ATLAS galaxies with $z_{\rm ph} \ge 1.5$ in the present sample and in that selected by B15 (GN12) using slightly more restrictive criteria. The effect of using different masks (the 2013 and 2015 \emph{Planck} masks) is also shown. The solid black line represents the auto-power spectra for the best-fit value of the bias parameter given in the inset (see also Table~\ref{b_A_maps_masks_old}) and for the redshift distribution of B15. The estimated bandpowers are plotted with an offset with respect to the bin centers for a better visualization. The error bars are computed using the analytical prescription ($\sqrt{\text{diag}(\text{Cov}^{gg})}$ with $\text{Cov}^{gg}$ given by Eq.~(\ref{eqn:covs}) and evaluated using the estimated bandpowers).   \label{fig:gg_data}}
\end{figure}

The exploitation of the 2015 CMB lensing map has the effect of shrinking the error bars, on average, by approximately 15\% with respect to the
previous data release due to the augmented  \emph{Planck} sensitivity. All shifts in the cross-power spectra based on the 2013 and on the 2015
releases are within $1\,\sigma$. As illustrated by Fig.~\ref{fig:gg_data}, the auto-power spectra, $C_{\ell}^{gg}$, of H-ATLAS galaxies in the present sample and in the B15 one are consistent with each other: differences are well within $1\,\sigma$. Table~\ref{b_A_maps_masks_old} shows that the various combinations of lensing maps, galaxy catalogues and masks we have considered in Fig.~\ref{fig:kg_data} lead to very similar values of the $A$ and $b$ parameters, \emph{if the redshift distribution of B15 is used}. Note that the errors on parameters given in Table~\ref{b_A_maps_masks_old} as well as in the following similar tables, are slightly smaller than those that could be inferred from the corresponding figures. This is because the errors on each parameter given in the tables are obtained marginalizing over the other parameter.

\subsection{Tomographic analysis}
\label{subsec:tomography}
As discussed in the previous sub-section, {if we use the redshift distribution of B15} the impact of the new \textit{Planck} convergence map and mask, and of the new H-ATLAS overdensity map on the $A$ and $b$ parameters is very low. However, significant differences are found using the new redshift distribution for $z_{\rm ph}\ge 1.5$ built in this paper and shown by the blue solid line in the bottom panel of Fig.~\ref{fig:dndz}. Compared to B15, the best fit value of the bias parameter increases from $b=2.80^{+0.12}_{-0.11}$  to $b=3.54^{+0.15}_{-0.14}$ and the cross-correlation amplitude decreases from  $A = 1.62 \pm 0.16$  to $A = 1.45^{+0.14}_{-0.13}$ (see Tables~\ref{b_A_maps_masks_old} and \ref{b_a_results}).

As in B15, we get a highly significant detection of the cross-correlation, at $A/\sigma_A\simeq 10.3\, \sigma$ and again a value of $A$ higher than the expected $A=1$ is indicated by the data.

\begin{figure} 
\plotone{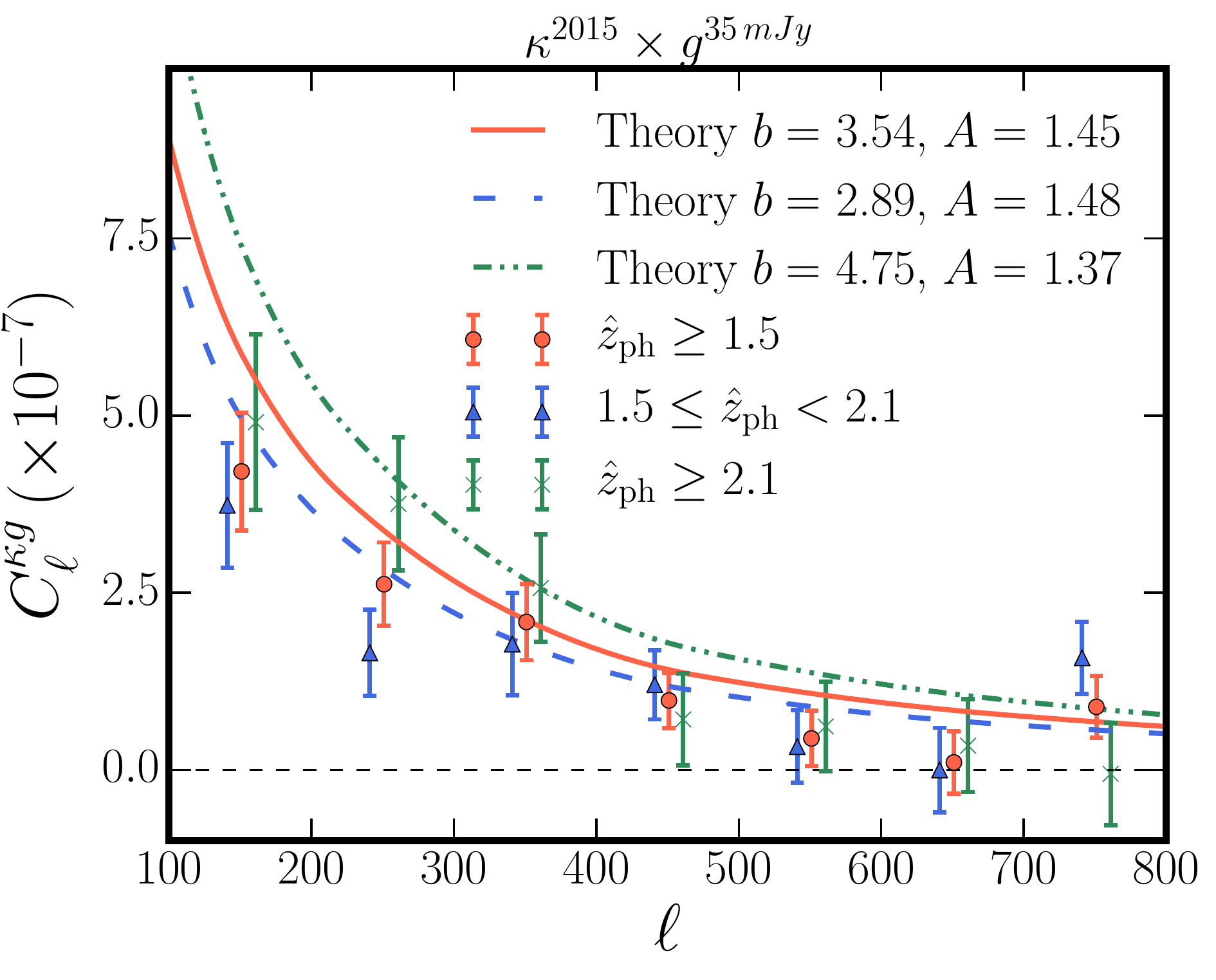}
\caption{Cross-power spectra between the 2015 \emph{Planck} CMB lensing map and the H-ATLAS galaxy sample for different redshift intervals: $z_{\rm ph} \ge 1.5$ (red circles), $1.5 \le z_{\rm ph} < 2.1$ (blue triangles), and $z_{\rm ph}\ge 2.1$ (green crosses). Uncertainties
are derived as for bandpowers in Fig.~\ref{fig:kg_data}. The red solid, blue dashed and green dot-dashed lines are the corresponding cross-power spectra for the best-fit bias and amplitude parameters obtained combining the data on the auto- and cross-power spectra (see Table~\ref{b_a_results}). The adopted redshift distributions are shown by the blue lines in Fig.~\ref{fig:dndz}.
\label{fig:kg_data_tomo}}
\end{figure}

Figure~\ref{fig:kg_data_tomo} shows the cross-correlation power spectrum for the 3 redshift intervals we have considered. The error bars were estimated with Monte Carlo simulations as described above. Their relative sizes scale, as expected, with the number of objects in each photo-$z$ interval, reported in Table~\ref{herschel_patches}. In all cases, the detection of the signal is highly significant. The chi-square value for the null hypothesis, i.e. no correlation between the two cosmic fields, computed taking into account bin-to-bin correlations, is $\chi_{\text{null}}^2 = \hat{\mathbf{C}}^{\kappa g}_{L} \,(\text{Cov}^{\kappa g}_{LL'})^{-1}\, \hat{\mathbf{C}}^{\kappa g}_{L'} \simeq 88$ for $\nu=7$ d.o.f., corresponding to a $\simeq 22\sigma$ rejection for the full sample ($z_{\rm ph} \ge 1.5$). For the $1.5 \le z_{\rm ph} < 2.1$ and $z_{\rm} \ge 2.1$ intervals we found $\chi^2_{\rm null} \simeq 47$ and $\chi^2_{\rm null} \simeq 64$, respectively, corresponding to  $10.7\sigma$ and $15\sigma$ rejections.

\begin{deluxetable}{cccc}
\tabletypesize{}
\tablecolumns{4}
\tablewidth{0pt}
\tablecaption{Comparison of the $\{b,A\}$ values obtained from the joint $\kappa g + gg$ analysis for the combinations of maps and
masks reported in Fig.~\ref{fig:kg_data} and adopting the redshift distribution of B15. \label{b_A_maps_masks_old}}
\tablehead{
\colhead{Datasets} & \colhead{Mask} &  \colhead{$b$} & \colhead{A} }
\startdata
$\kappa^{2013} \times  g^{GN12}$ & 2013  & $2.80^{+0.12}_{-0.11}$ & $1.62^{+0.16}_{-0.16}$ \\
$\kappa^{2013} \times  g^{GN12}$ & 2015  & $2.86^{+0.12}_{-0.12}$ & $1.68^{+0.19}_{-0.19}$ \\
$\kappa^{2015} \times g^{GN12}$ & 2015  & $2.85^{+0.12}_{-0.12}$ & $1.61^{+0.16}_{-0.16}$ \\
$\kappa^{2015} \times g^{35\rm{mJy}}$ & 2015  &$2.79^{+0.12}_{-0.12}$ & $1.65^{+0.16}_{-0.16}$ \\
\enddata
\tablenotetext{a}{The analysis is performed on the $z_{\rm ph} \ge 1.5$ sample for consistency with B15.}
\end{deluxetable}

\begin{figure} 
\plotone{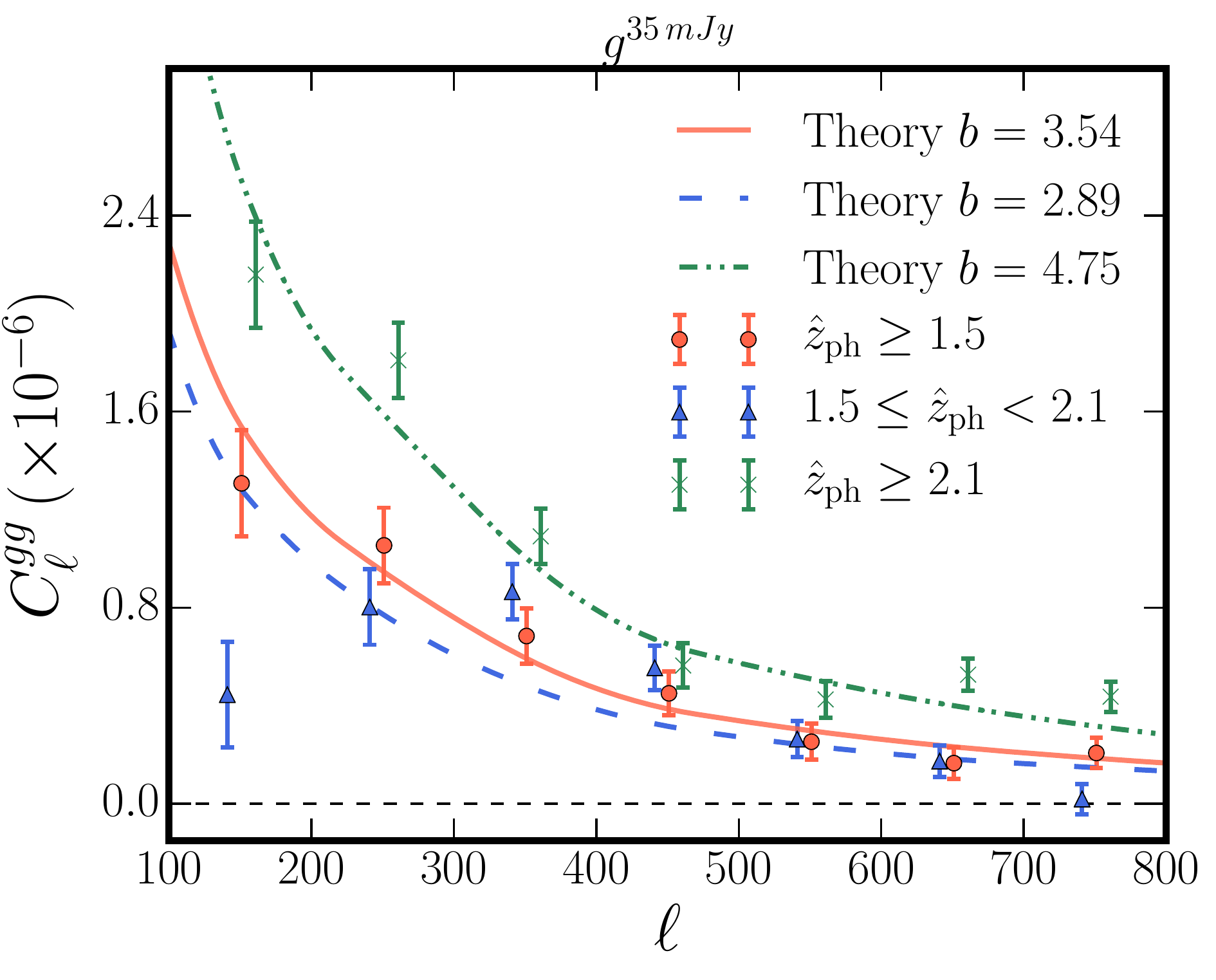}
\caption{H-ATLAS galaxy auto-power spectra in different redshift intervals: $z_{\rm ph} \ge 1.5$ (red circles), $1.5 \le z_{\rm ph} < 2.1$
(blue triangles), and $z_{\rm ph}\ge 2.1$ (green crosses). Uncertainties are derived as for bandpowers in Fig.~\ref{fig:gg_data}. The red solid, blue dashed and green dot-dashed lines are the galaxy auto-power spectra for the \emph{Planck} cosmology and the best-fit bias and amplitude found for the $z \ge 1.5$, $1.5 \le z < 2.1$, and $z \ge 2.1$ photo-$z$ bins respectively. The theory lines refer to the $dN/dz$ built in this paper and also used in Fig.~\ref{fig:kg_data_tomo}. \label{fig:gg_data_tomo}}
\end{figure}

There is  a hint of a stronger cross-correlation signal for the higher redshift interval. The indication is however weak. A much stronger indication of an evolution of the clustering properties (increase with redshift of the bias factor) of galaxies is apparent in Fig.~\ref{fig:gg_data_tomo} and in Table~\ref{b_a_results}. However the auto--power spectrum for the $1.5 \le z_{\rm ph} < 2.1$ interval shows a puzzling lack of power in the first multipole bin. This feature, not observed in the cross-power spectrum for the same photo-$z$ bin, may be due to systematic errors in the photometric redshift estimate.

\begin{deluxetable}{ccccc}
\tabletypesize{}
\tablecolumns{5}
\tablewidth{0pt}
\tablecaption{Linear bias and cross-correlation amplitude as determined using jointly the reconstructed galaxy auto- and cross-spectra in the different redshift bins. \label{b_a_results}}
\tablehead{
\colhead{Bin} & \colhead{$b$} & \colhead{A} & \colhead{$\chi^2$/d.o.f.} & \colhead{p-value}}
\startdata
\multicolumn{5}{c}{Diagonal covariance matrices approximation (Eq.~\ref{eqn:covs})}\\
\cline{1-5} \\
$z_{\rm ph} \ge 1.5$  & $3.54^{+0.15}_{-0.14}$  &  $1.45^{+0.14}_{-0.13}$ & $10.6/12$ & $0.56$ \\
$1.5 \le z_{\rm ph} < 2.1$  & $2.89^{+0.23}_{-0.23}$   &  $1.48^{+0.20}_{-0.19}$  & $29.5/12$ & $0.003$ \\
$z_{\rm ph} \ge 2.1$  & $4.75^{+0.24}_{-0.25}$   &  $1.37^{+0.16}_{-0.16}$ & $9.6/12$ & $0.65$ \\
\cline{1-5} \\
\multicolumn{5}{c}{Non-diagonal covariance matrices approximation (Eq.~21 of B15)}\\
\cline{1-5} \\
$z_{\rm ph} \ge 1.5$  & $3.53^{+0.15}_{-0.15}$  &  $1.45^{+0.14}_{-0.13}$ & $8.75/12$ & $0.72$ \\
$1.5 \le z_{\rm ph} < 2.1$  & $2.88^{+0.23}_{-0.25}$   &  $1.48^{+0.20}_{-0.19}$  & $23.1/12$ & $0.03$ \\
$z_{\rm ph} \ge 2.1$  & $4.74^{+0.24}_{-0.24}$   &  $1.36^{+0.16}_{-0.16}$ & $8.5/12$ & $0.75$ \\
\cline{1-5} \\
\multicolumn{5}{c}{Covariance matrices from MC simulations}\\
\cline{1-5} \\
$z_{\rm ph} \ge 1.5$  & $3.56^{+0.17}_{-0.17}$  &  $1.39^{+0.15}_{-0.14}$ & $8.37/12$ & $0.76$ \\
$1.5 \le z_{\rm ph} < 2.1$  & $2.91^{+0.24}_{-0.24}$   &  $1.48^{+0.22}_{-0.21}$  & $33.7/12$ & $7.5 \times 10^{-4}$ \\
$z_{\rm ph} \ge 2.1$  & $4.80^{+0.25}_{-0.25}$   &  $1.37^{+0.17}_{-0.17}$ & $14.5/12$ & $0.27$ \\
\enddata
\tablenotetext{a}{The redshift distributions derived in this paper and shown by the black lines in Fig.~\ref{fig:dndz} were adopted. The best-fit values and $1\,\sigma$ errors are evaluated respectively as the $50$th, $16$th, and $84$th percentiles of the posterior distributions. The $\chi^2$'s are computed at the best-fit values. The results obtained including off-diagonal terms of the covariance matrices and using covariances based on simulations are also shown for comparison.}
\end{deluxetable}

\begin{figure} 
\plotone{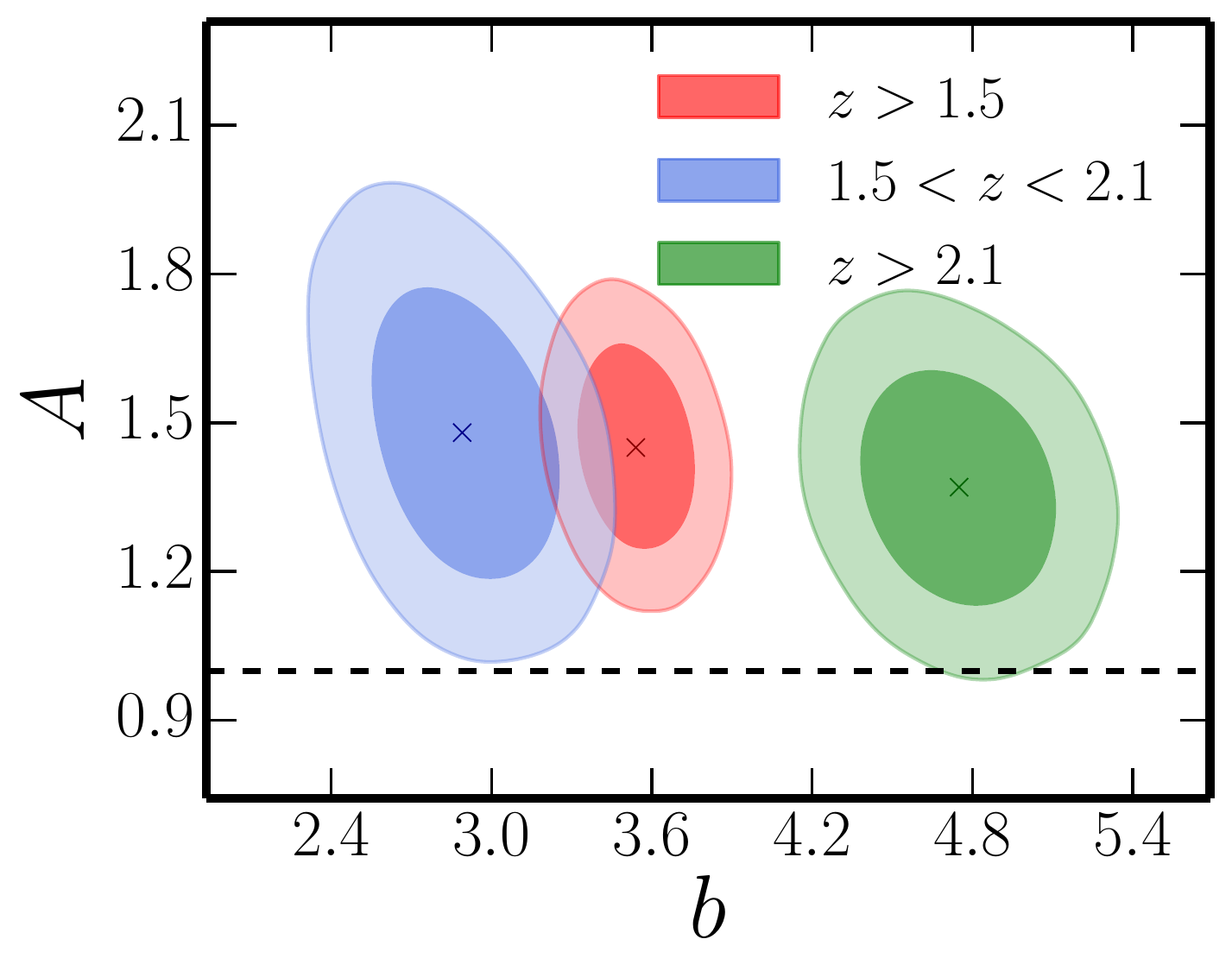}
\caption{Posterior distributions in the $(b,A)$ plane with the $68\%$ and $95\%$ confidence regions (darker and lighter colors respectively) for the three redshift intervals: $z_{\rm ph} \ge 1.5$ (red contours), $1.5 \le z_{\rm ph} < 2.1$ (blue contours), and $z_{\rm ph} \ge 2.1$ (green contours). The dashed line corresponds to the expected amplitude value $A=1$ (a magnification bias parameter $\alpha=3$ is assumed). The colored crosses mark the best-fit values reported in Table~\ref{b_a_results} for the three photo-z intervals.    \label{fig:b_A_kggg_2015}}
\end{figure}

The 68\% and 95\% confidence regions for the amplitude $A$ and the bias $b$, obtained from their posterior distributions combining the data on auto- and cross-spectra, are shown in Fig.~\ref{fig:b_A_kggg_2015}. We have $b=2.89\pm0.23$ and $A=1.48^{+0.20}_{-0.19}$ for the lower redshift bin and $b=4.75^{+0.24}_{-0.25}$ and  $A=1.37\pm 0.16$ for the higher-$z$ one (see Table~\ref{b_a_results}). The reduced $\chi^2$ associated with the best-fit values are close to unity, suggesting the consistency of the results, except for the $1.5 \le z_{\rm ph} < 2.1$ interval for which there is a large contribution to the $\chi^2$ from the auto-spectrum for the first multipole bin. In order to test the stability of the results with respect to the chosen covariance matrices estimation method, we redo the analysis with the non-diagonal approximation of B15 and the full covariance matrices from simulations: results are reported in Table~\ref{b_a_results}. As can be seen, in the former case the inclusion of non-diagonal terms induced by mode-coupling results in negligible differences with respect to our baseline analysis scheme. In the latter case we observe a rather small broadening of the credibility contours (dependent on the $z$-bin), from 2\% to 17\% for $b$ and from 6\% to 10\% for the cross-correlation amplitude $A$, with the biggest differences reported for the baseline $z \ge 1.5$ bin. The central value of $A$ for the baseline redshift bin is diminished by approximately 4\% even though $A > 1$ at $\gtrsim 2\sigma$. However, one might argue that the limited number of the available Planck CMB lensing simulations imposes limitations to the covariance matrices convergence. Given the stability of the results, we therefore adopt the diagonal approximation of Eq.~\ref{eqn:covs} as our baseline covariance matrices estimation method.

\subsection{Cross-correlation of galaxies in different redshift intervals}
Both the auto- and the cross-power spectra depend on the assumed redshift distribution; hence the inferred values of the (constant) bias and of the amplitude are contingent on it. A test of the reliability of our estimates can be obtained from the cross-correlation $C_{\ell}^{g_1g_2}$ between positions of galaxies in the lower redshift interval, $1.5 \le z_{\rm ph} < 2.1$ (indexed by subscript 1), with those in the higher redshift interval, $z_{\rm ph}\ge 2.1$ (subscript 2). Assuming, as we did in Eq.~(\ref{eqn:wg}), that the observed galaxy density fluctuations can be written as the sum of a clustering term with a magnification bias one as $\delta^{\rm obs}(\nver) = \delta^{\rm cl}(\nver) + \delta^{\mu}(\nver)$,  the cross-correlation among galaxies in the two intervals can be decomposed into four terms:
\begin{equation}
\label{eq:gg_cross}
C_{\ell}^{g_1g_2} = C_{\ell}^{\rm cl_1 cl_2} + C_{\ell}^{\rm cl_1 \mu_2} + C_{\ell}^{\rm \mu_1 cl_2} + C_{\ell}^{\rm \mu_1 \mu_2}.
\end{equation}
The first term results from the intrinsic correlations of the galaxies of the two samples and it is due to the overlap between the two redshift
distributions: if the two galaxy samples are separated in redshift, this term vanishes. The second term results from the lensing of
background galaxies due to the matter distribution traced by the low-$z$ sample galaxies, while the third one is related to the
opposite scenario: again, it is non-zero only if there is an overlap between the two $dN/dz$. The fourth term results from lensing induced
by dark matter in front of both galaxy samples. The relative amplitude of these terms, compared to the observed galaxy cross-power spectrum,
can provide useful insights on uncertainties in the redshift distributions.

\begin{figure}
\plotone{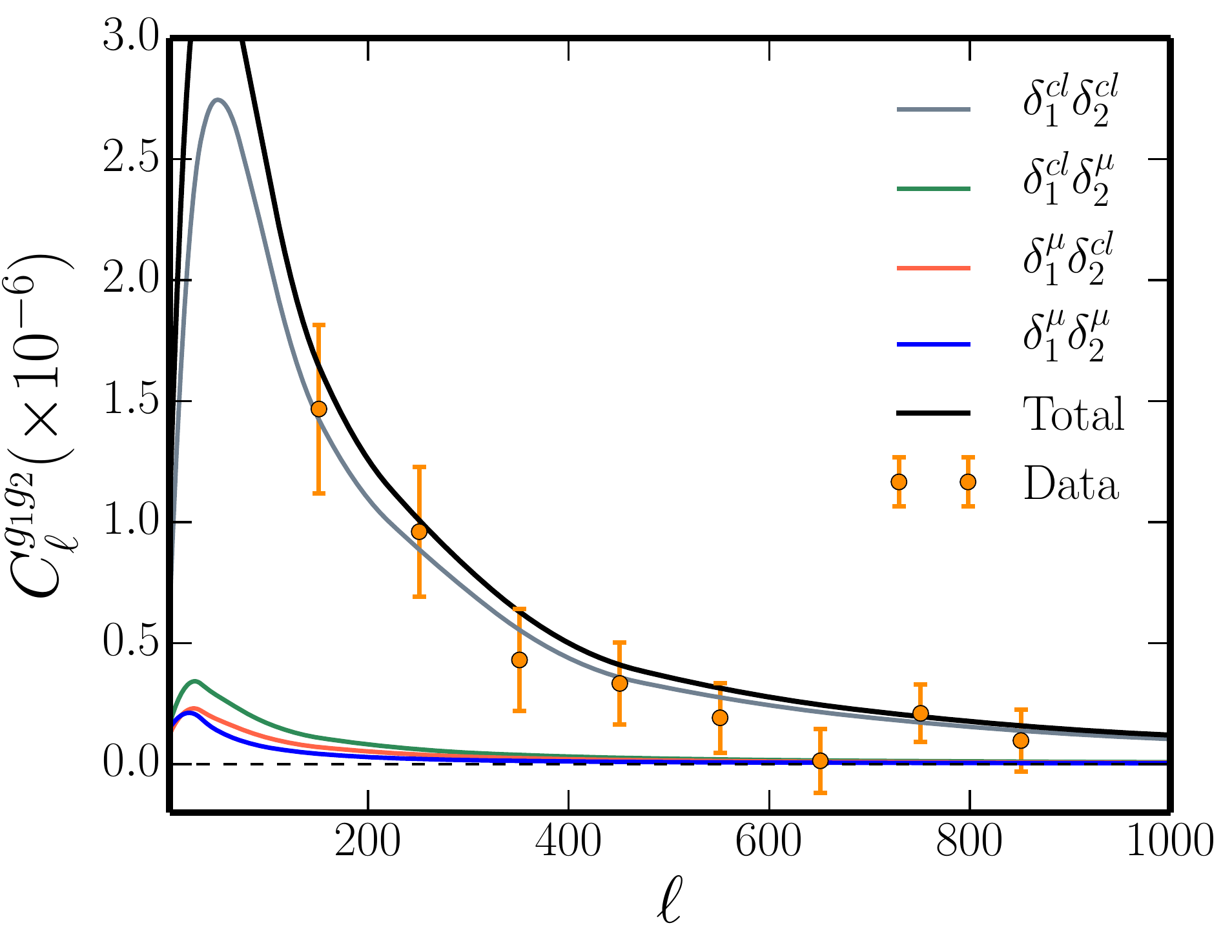}
\caption{Cross-correlation of angular positions between galaxies in the low-$z$ and in the high-$z$ redshift interval. The solid lines show the expected contributions from the various terms appearing in Eq.~(\ref{eq:gg_cross}). Note that the ``Total'' line is \emph{not} a fit to the data.
\label{fig:gg_cross}}
\end{figure}

The measured $\hat{C}^{g_1g_2}_{\ell}$ is shown in Fig.~\ref{fig:gg_cross}. The expected contributions of the four aforementioned terms are
computed using the bias values reported in Table~\ref{b_a_results}, and the redshift distributions shown in Fig.~\ref{fig:dndz}. We remind that the assumed value for the rms uncertainty is $\sigma_{\Delta z/(1+z)}=0.26$. The figure shows that the expected amplitude of the intrinsic correlation term is dominant with respect to the magnification bias related ones and that the observed signal is in good agreement with expectations. No signs of inconsistencies affecting redshift distributions are apparent.

\subsection{Effect of different choices for the SED}
\label{sec:SED}
The assumed SED plays a key role in the context of template fitting approaches aimed at photo-$z$ estimation. It is then crucial to test the robustness of the results presented here against different choices for it. To this end we constructed a catalogue with photo-$z$ estimates based on the best fitting SED template of \cite{pearson13} and applied the full analysis pipeline described in Sect.~\ref{sec:analysis}.

\begin{figure}
\plotone{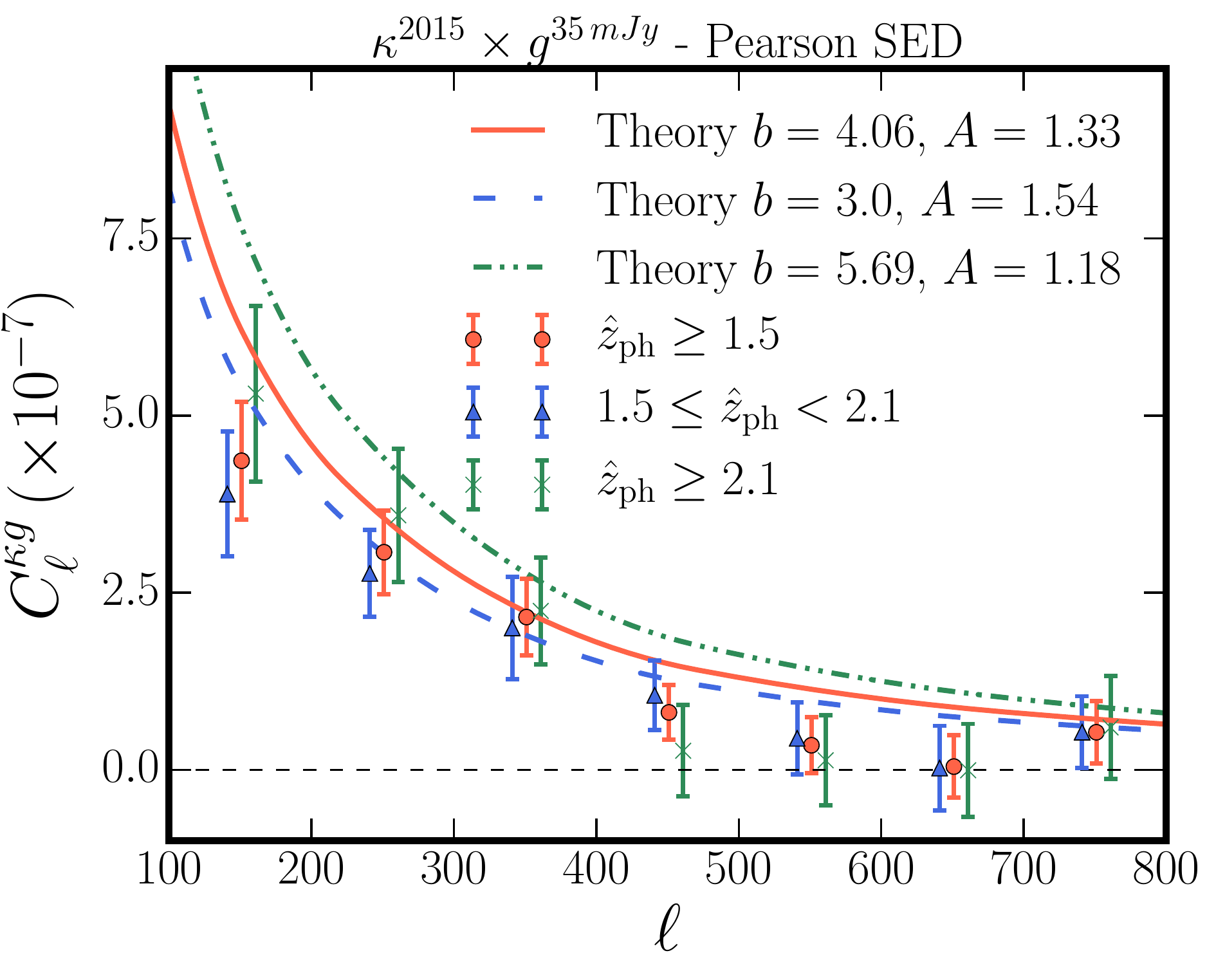}
\caption{Cross-power spectra between the 2015 \emph{Planck} CMB lensing map and the H-ATLAS galaxy sample built with the SED of \cite{pearson13} for different redshift intervals: $z_{\rm ph} \ge 1.5$ (red circles), $1.5 \le z_{\rm ph} < 2.1$ (blue triangles), and $z_{\rm ph}\ge 2.1$ (green crosses). Uncertainties are derived as for bandpowers in Fig.~\ref{fig:kg_data}. The red solid, blue dashed and green dot-dashed lines are the corresponding cross-power spectra for the best-fit bias and amplitude parameters obtained combining the data on the auto- and cross-power spectra (see Table~\ref{b_a_results}). The adopted redshift distributions are shown by the orange lines in Fig.~\ref{fig:dndz}.
\label{fig:kg_data_tomo_Pearson}}
\end{figure}

\begin{figure} 
\plotone{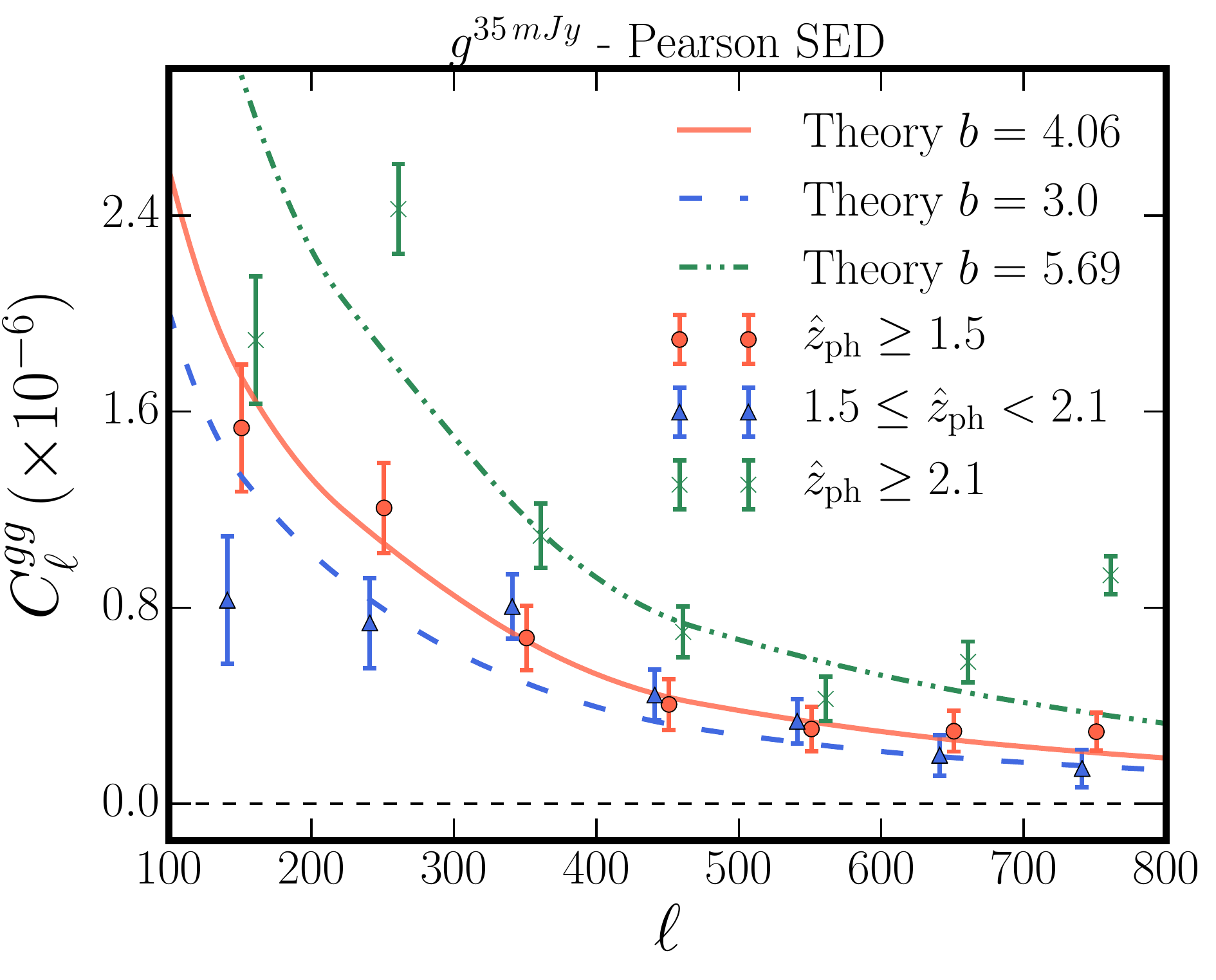}
\caption{H-ATLAS galaxy auto-power spectra in different redshift intervals: $z_{\rm ph} \ge 1.5$ (red circles), $1.5 \le z_{\rm ph} < 2.1$ (blue triangles), and $z_{\rm ph}\ge 2.1$ (green crosses). The SED template of \cite{pearson13} was adopted to estimate photo-$z$. Uncertainties are derived as for bandpowers in Fig.~\ref{fig:gg_data}. The red solid, blue dashed and green dot-dashed lines are the galaxy auto-power spectra for the \emph{Planck} cosmology and the best-fit bias and amplitude found for the $z \ge 1.5$, $1.5 \le z < 2.1$, and $z \ge 2.1$ photo-$z$ bins respectively. The theory lines refer to the $dN/dz$ built in this paper and also used in Fig.~\ref{fig:kg_data_tomo_Pearson}. \label{fig:gg_data_tomo_Pearson}}
\end{figure}

\begin{figure}
\plotone{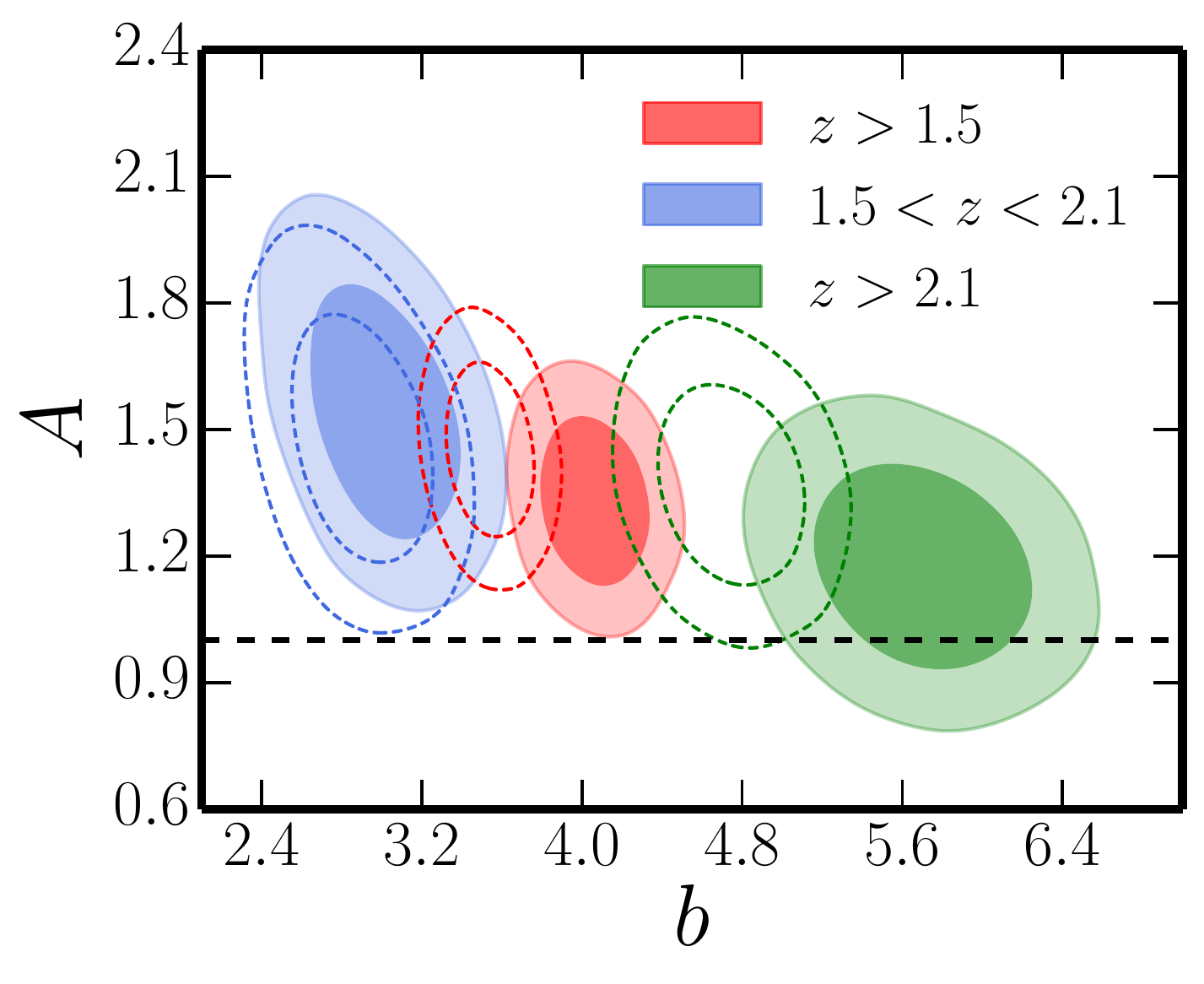}
\caption{Posterior distributions in the $(b,A)$ plane with the $68\%$ and $95\%$ confidence regions (darker and lighter colors respectively) plane based on the $dN/dz$ obtained using the \citet{pearson13} best fitting template (filled contours) and using the baseline SMM~J2135-0102 SED (dashed contours) for the three redshift intervals: $z_{\rm ph} \ge 1.5$ (red contours), $1.5 \le z_{\rm ph} < 2.1$ (blue contours), and $z_{\rm ph} \ge 2.1$ (green contours).
\label{fig:b_A_SMM_Pearson}}
\end{figure}

The cross- and auto-power spectra extracted adopting the SED template of \cite{pearson13} are shown in Fig.~\ref{fig:kg_data_tomo_Pearson} and \ref{fig:gg_data_tomo_Pearson} respectively.
In Fig.~\ref{fig:b_A_SMM_Pearson} we compare the credibility regions for the bias $b$ and cross-correlation amplitude $A$ obtained with the $dN/dz$ based on the \citet{pearson13} best fitting template (filled contours) with that based on the baseline SMM~J2135-0102 SED in the three photo-$z$ intervals. The best fit parameter values for the \citet{pearson13} SED are reported in Table~\ref{b_a_results_Pearson}.

The \citet{pearson13} SED leads to a cross-correlation amplitude consistent with SMM~J2135-0102--based results for the $1.5 \le z_{\rm ph} < 2.1$ interval and for the full $z_{\rm ph}\ge 1.5$ sample, although the deviation from $A=1$ has a slightly lower significance level: we have $A>1$ at $\simeq 2.5\,\sigma$ (it was $\simeq 3.5\,\sigma$ in the SMM~J2135-0102 case). For the high-$z$ bin we get consistency with $A=1$ at the $\simeq 1\,\sigma$ level. Also, there no longer a lack of power in the first multipole bin of the galaxy auto-power spectrum for the lower-$z$ interval. The shifts in the $A$ parameter values are associated to changes in the bias value: as we move toward higher redshifts, the bias parameter grows increasingly larger compared to that found using the  SMM~J2135-0102 SED. Adopting an effective redshift $z_{\rm eff}=2.15$ for the high-$z$ sample we find that the best fit value $b=5.69$ corresponds to a characteristic halo mass $\log(M_{\rm H}/M_\odot)=13.5$, substantially larger than found by other studies \citep{Xia2012,Hickox2012,Cai2013,Viero2013,Hildebrandt2013} to the point of being probably unrealistic. 

\subsection{Redshift dependence of the galaxy bias}
\label{sec:bz}
Using a single, mass independent, bias factor throughout each redshift interval is certainly an approximation, although the derived estimates can be interpreted as effective values. In fact it is known \citep[e.g., ][]{sheth2001,mo10} that the bias function increases rapidly with the halo mass, $M_H$, and with $z$ at fixed $M_H$.

To test the stability of the derived cross-correlation amplitude $A$ against a more refined treatment of the bias parameter we have worked out an estimate of the expected effective bias function, $b_0(z)$, for our galaxy population. We started from the linear halo bias model $b(M_H; z)$ computed via the excursion set approach \citep{lapi14a}. The halo mass distribution was inferred from the observationally determined, redshift dependent, luminosity function, $N(\log{L_{\rm SFR};z})$, where $L_{\rm SFR}$ is the total luminosity produced by newly formed stars, i.e. proportional to the Star Formation Rate (SFR). To this end we exploited the relationship between $L_{\rm SFR}$ and $M_H$ derived by \citet{aversa15} by means of the abundance matching technique. Finally, we computed the luminosity-weighted bias factor as a function of redshift
\begin{equation}
b_0(z) = \frac{\int d\log{L_{\rm SFR}}N(\log{L_{\rm SFR};z}) b(L_{\rm SFR}; z)}{\int d\log{L_{\rm SFR}}N(\log{L_{\rm SFR};z})},
\end{equation}
where the integration runs above $L_{\rm min}(z)$, the luminosity associated to our flux density limit $S_{250}=35\,$mJy at 250 $\mu$m.

To quantify deviations, requested by the data,  from the expected effective bias function, $b_0(z)$, we have introduced a scaling parameter $\mathcal{A}_{\rm bias}$ so that the effective bias function used in the definition of the galaxy kernel $W^{g}(z)$ [eq.~(\ref{eqn:wg})] is $b(z) = \mathcal{A}_{\rm bias}b_0(z)$.

\begin{figure} %8
\plotone{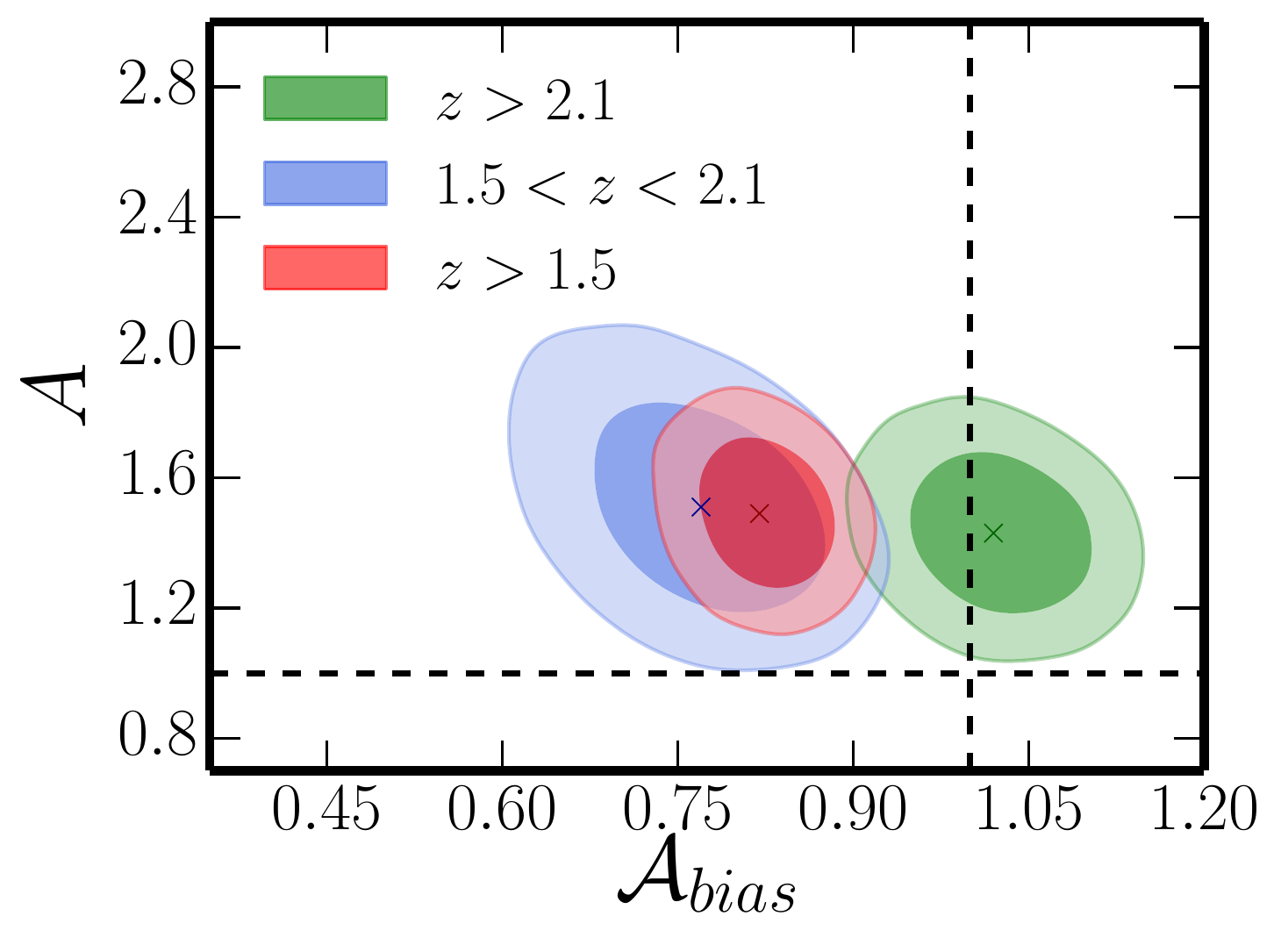}
\caption{Posterior distributions in the $(\mathcal{A}_{\rm bias},A)$ plane with the $68\%$ and $95\%$ confidence regions (darker and lighter colors respectively) for the three redshift intervals: $z_{\rm ph} \ge 1.5$ (red contours), $1.5 \le z_{\rm ph} < 2.1$ (blue contours), and $z_{\rm ph} \ge 2.1$ (green contours). The dashed lines correspond to $A=1$ and $\mathcal{A}_{\rm bias}=1$. The colored crosses mark the best-fit values reported in Table~\ref{A_bias_A_results}. \label{fig:A_bias_A_kggg_2015}}
\end{figure}

\begin{figure} %9
\plotone{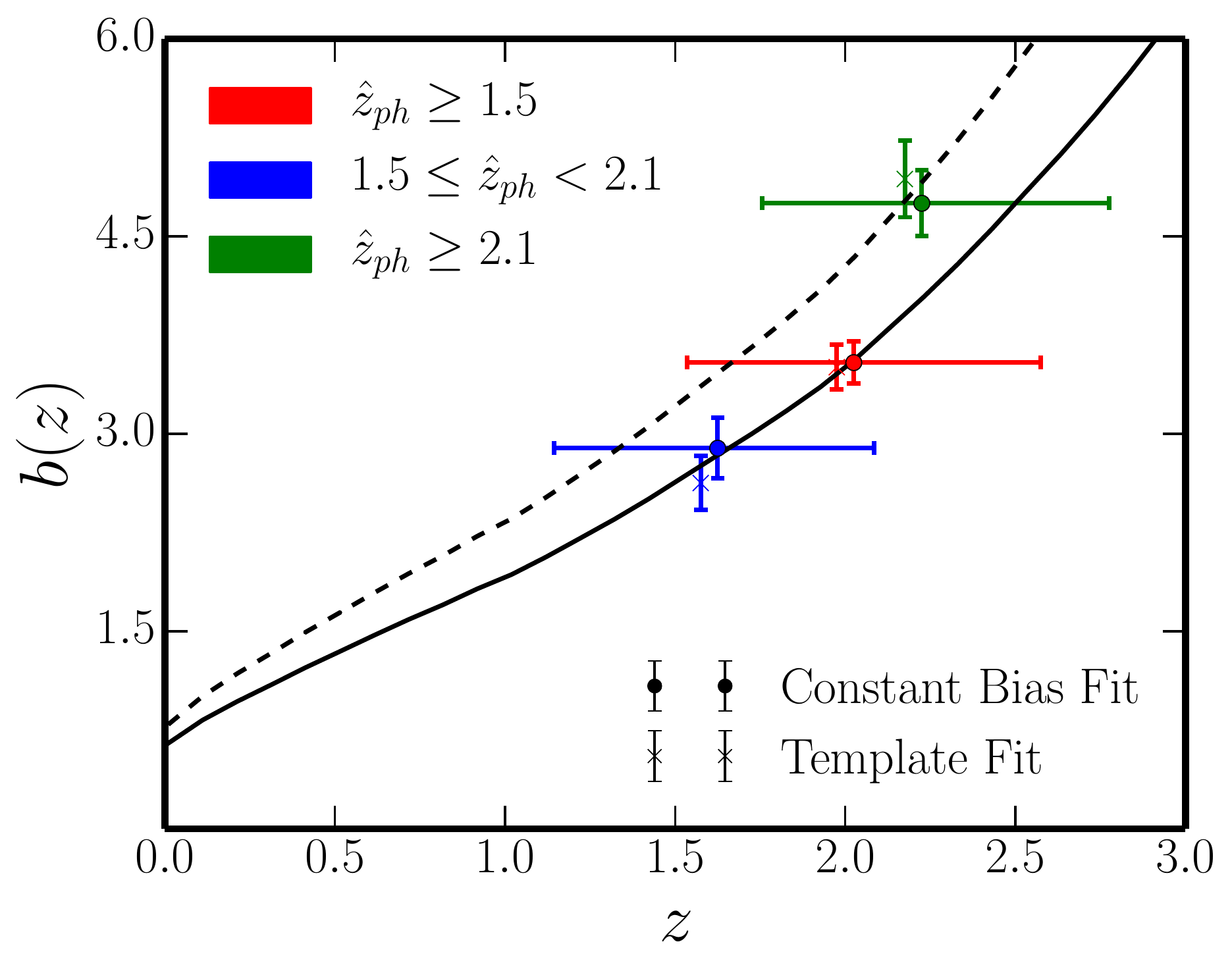}
\caption{Effective bias functions. The dashed line corresponds to $b_0(z)$, while the solid line shows $b(z)$ with $\mathcal{A}_{\rm bias}=0.82$, the best fit value for $z_{\rm ph} \ge 1.5$. The data points show the best fit values of the bias parameter at the median redshifts of the distributions for $z_{\rm ph} \ge 1.5$ (red), $1.5 \le z_{\rm ph} < 2.1$ (blue) and $z_{\rm ph}\ge 2.1$ (green). In the ``Template Fit'' case $b(z) = \mathcal{A}_{\rm bias}b_0(z)$. Horizontal error bars indicate the $z$-range that includes 68\% of each of the redshift distribution.  \label{fig:b_z_comp}}
\end{figure}

The 68\% and 95\% confidence regions in the ($\mathcal{A}_{\rm bias}$,$A$) plane are shown in Fig.~\ref{fig:A_bias_A_kggg_2015} and the central values of the posterior distributions are reported in Table~\ref{A_bias_A_results}, while the corresponding bias evolution is shown in Fig.~\ref{fig:b_z_comp}. On one side we note that $\mathcal{A}_{\rm bias}$ is found to be not far from unity, indicating that our approach to estimate the effective bias function is reasonably realistic. The largest deviations of $\mathcal{A}_{\rm bias}$ from unity happen for the lower redshift interval that may be more liable to errors in photometric redshift estimates. However, the values of the cross-correlation amplitude $A$ are in agreement with the previous results of Table~\ref{b_a_results}, showing that our constant bias approach does not significantly undermines the derived value of $A$.

\begin{deluxetable}{ccccc}
\tabletypesize{}
\tablecolumns{5}
\tablewidth{0pt}
\tablecaption{Same as Table~\ref{b_a_results} but for analysis based on SED template of \cite{pearson13}.  \label{b_a_results_Pearson}}
\tablehead{
\colhead{Bin} & \colhead{$b$} & \colhead{A} }
\startdata
$z_{\rm ph} \ge 1.5$  & $4.06^{+0.18}_{-0.18}$  &  $1.33^{+0.13}_{-0.13}$ \\
$1.5 \le z_{\rm ph} < 2.1$  & $3.00^{+0.24}_{-0.25}$   &  $1.54^{+0.20}_{-0.19}$  \\
$z_{\rm ph} \ge 2.1$  & $5.69^{+0.36}_{-0.36}$   &  $1.18^{+0.16}_{-0.16}$ \\
\enddata
\tablenotetext{a}{The redshift distributions derived in this paper and shown by the orange lines in Fig.~\ref{fig:dndz} were adopted.}
\end{deluxetable}

\subsection{Results dependence on flux limit}
\label{sec:flux_test}
\begin{figure}
\plotone{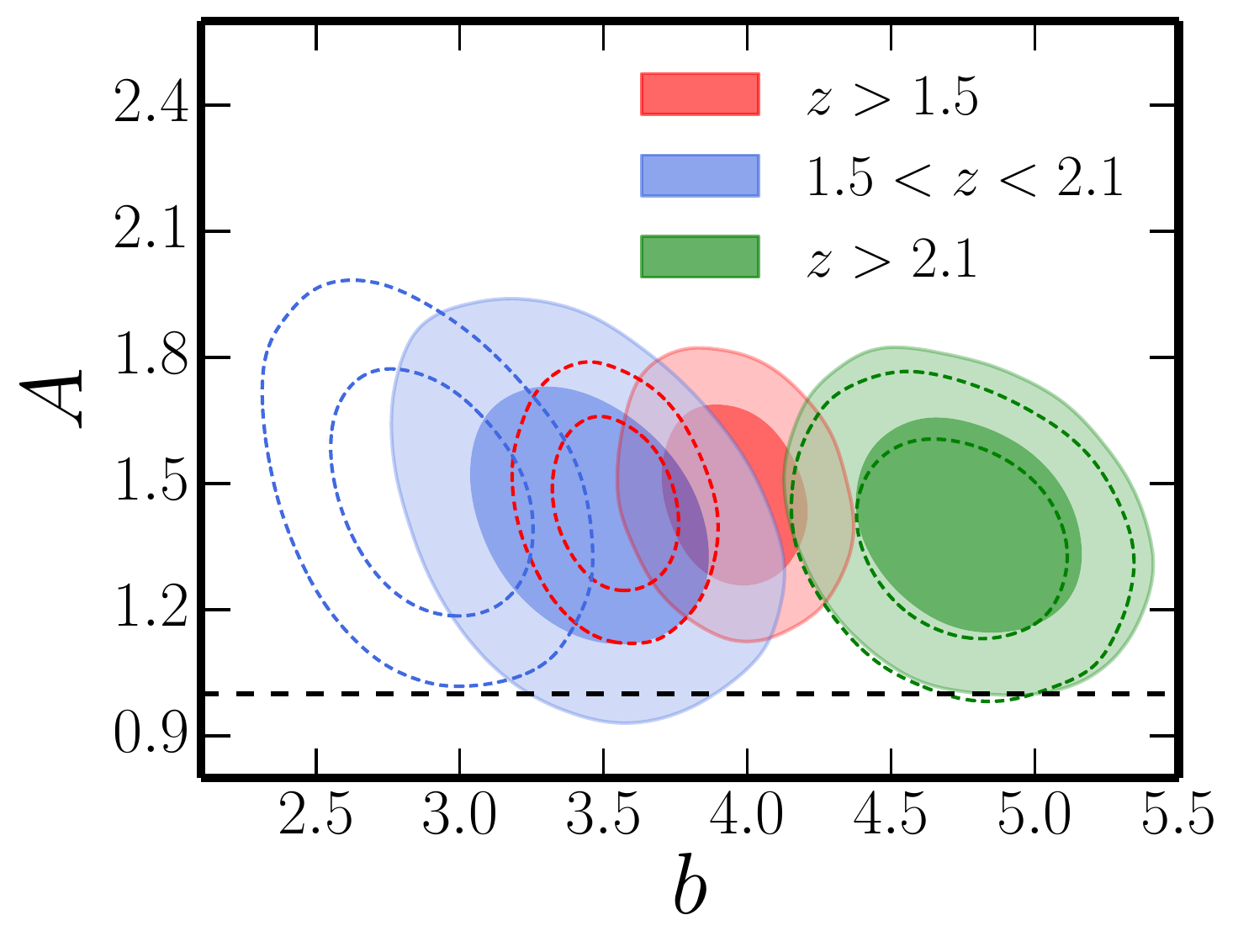}
\caption{Posterior distributions in the $(b,A)$ plane obtained requiring a $\ge 5\sigma$ detection at $350\,\mu$m (solid contours) compared with distributions obtained with our baseline selection criterion ($\ge 3\sigma$ detection; dashed contours) for the three redshift intervals: $z_{\rm ph} \ge 1.5$ (red contours), $1.5 \le z_{\rm ph} < 2.1$ (blue contours), and $z_{\rm ph} \ge 2.1$ (green contours).
\label{fig:b_A_SMM_3sigma_5sigma}}
\end{figure}

To check the stability of the results against changes in the selection criterion (ii) formulated in Sec.~\ref{subsec:herschel} we built a new catalogue with objects complying with criteria (i), (iii) and with a (ii.b) $\ge 5\sigma$ detection at $350\,\mu$m, and applied the pipeline outlined in Sec.~\ref{sec:analysis} in the three photo-$z$ intervals. Raising the detection threshold at 350\,$\mu$m has the effect of decreasing the statistical errors on photometric redshifts because of the higher signal-to-noise photometry and of favoring the selection of redder, higher-$z$ galaxies; the total number of sources decreases by approximately 20\%. The credibility regions in the $(b,A)$ plane are presented in Fig.~\ref{fig:b_A_SMM_3sigma_5sigma} while the best fit values of the parameters values are reported in Table~\ref{b_a_results_flux}.

The inferred cross-correlation amplitudes are consistent with the previous estimates within the statistical error in all of the three photo-$z$ bins ($A>1$ at $\sim2-3\,\sigma$). The value of bias parameter for the low-$z$ bin increases (dragging also the value for the full $z_{\rm ph}\ge1.5$ sample), while the value for the high-$z$ interval is essentially unchanged. This is likely due to the fact that by requiring at least a $5\sigma$ detection at 350 $\mu$m, we select objects which are intrinsically more luminous, hence more biased. The high-$z$ sample is not affected by the higher threshold because at such distances we already detect only the most luminous objects (Malmquist bias). At the power spectrum level we find that, for both the total $z_{\rm ph}\ge1.5$ sample and the low-$z$ sample, the cross-power spectra are less affected by the modification of the selection criteria, while the galaxy auto-power spectra are systematically above those obtained with the $3\sigma$ selection at $350\,\mu$m. Errors in the photo-$z$ estimates may also have a role, particularly for the low-$z$ sample; a hint in this direction is that the lack of power of $C_{\ell}^{gg}$ in the lowest multipole bin for the low-$z$ sample is no longer present in the case of the $5\,\sigma$ selection.

\begin{deluxetable}{ccccc}
\tabletypesize{}
\tablecolumns{5}
\tablewidth{0pt}
\tablecaption{Best fit values of the cross-correlation, $A$, and bias, $\mathcal{A}_{\rm bias}$, amplitudes obtained combining the observed $\kappa g$ and $gg$ spectra. \label{A_bias_A_results}}
\tablehead{
\colhead{Bin} & \colhead{$\mathcal{A}_{\rm bias}$} & \colhead{A} & \colhead{$\chi^2$/ d.o.f.} & \colhead{p-value}}

\startdata
$z_{\rm ph} \ge 1.5$  & $0.82^{+0.04}_{-0.04}$  &  $1.49^{+0.15}_{-0.15}$ & $9.5/12$ & $0.66$ \\
$1.5 \le z_{\rm ph} < 2.1$  & $0.77^{+0.06}_{-0.07}$   &  $1.51^{+0.22}_{-0.20}$  & $25.7/12$ & $0.01$ \\
$z_{\rm ph} \ge 2.1$  & $1.02^{+0.05}_{-0.05}$   &  $1.43^{+0.16}_{-0.16}$ & $9.6/12$ & $0.65$ \\
\enddata
\tablenotetext{a}{ The reduced $\chi^2$ are computed at the best-fit values.}
\end{deluxetable}

\subsection{Other tests}
The bias parameter is also influenced by non-linear processes at work on small scales. Thus it can exhibit a scale dependence. At an effective redshift of $z_{\rm eff}\simeq 2$ the multipole range $100 < \ell < 800$ corresponds to physical scales of $\approx 50 - 6$ Mpc (or $k \approx 0.03-0.2\, h/$Mpc). Moreover, \emph{Planck} team does not include multipoles $\ell>400$ in cosmological analysis based on the auto-power spectrum due do to some failed curl-mode tests. We have repeated the MCMC analysis restricting both the cross- and auto-power spectra to $\ell_{\rm max} = 400$ and found $b=3.58\pm0.18$ and $A = 1.47\pm0.14$ for the baseline photo-$z$ bin, fully consistent with the numbers shown in Table~\ref{b_a_results}. For the low-$z$ bin we obtained  $b=2.76\pm0.28$ and $A = 1.46\pm0.22$, while for the high-$z$ one we found $b=4.81\pm0.30$ and $A = 1.45\pm0.17$. Therefore it looks unlikely that the higher than expected value of $A$ can be 
ascribed to having neglected non-linear effects, to a scale-dependent bias or to issues associated with the \emph{Planck} lensing map.

To check the effect of our choice of the background cosmological parameters we have repeated the analysis adopting the $\hbox{WMAP}9 + \hbox{SPT} + \hbox{ACT} + \hbox{BAO} + H_0$ ones \citep{wmap9}. Both $A$ and $b$ changed by $<5\%$.

The values of the bias parameter are stable and well-constrained in all redshift intervals and can therefore be exploited to gain information on the effective halo masses and SFRs of galaxies. Using the relations obtained by \cite{aversa15} one can relate the galaxy luminosities to the SFRs and to the dark matter halo masses, $M_H$. The results are reported in Table~\ref{SFR_results}. The SFRs are a factor of several above the average main sequence values \citep[see][]{rodighiero14,speagle14}. The host halo masses suggest that these objects constitute the progenitors of local massive spheroidal galaxies  \citep[see][]{lapi11,lapi14b}).

Temperature-based reconstruction of the CMB lensing signal may be contaminated by a number of foregrounds such as the thermal Sunyaev-Zel'dovich effect and extragalactic sources. Of particular concern for the present analysis is the possibility of the CIB emission leakage into the lensing map through the temperature maps used for the lensing estimation, as it  strongly correlates with the CMB lensing signal \citep{planck_cib:2013}. The H-ATLAS galaxies are well below the Planck detection limits (their flux densities at 148GHz are expected to be in the range $0.1 - 1$ mJy, hence are much fainter than sources masked by \cite{PlanckCollaborationXV2015}), thus they are part of the CIB measured by Planck.  Foreground induced biases to CMB lensing reconstruction have been extensively studied by \cite{vanEngelen:2013} and \cite{osborne:2013}. These authors concluded that the impact of these sources of systematic errors should be small due to \emph{Planck}'s resolution and noise levels.

\begin{deluxetable}{ccccc}
\tabletypesize{}
\tablecolumns{5}
\tablewidth{0pt}
\tablecaption{Best fit values of the cross-correlation amplitudes $A$ and galaxy linear bias $b$ obtained requiring a $\ge 5\sigma$ detection at $350\,\mu$m and combining the observed $\kappa g$ and $gg$ spectra. \label{b_a_results_flux}}
\tablehead{
\colhead{Bin} & \colhead{$b$} & \colhead{A} }
\startdata
$z_{\rm ph} \ge 1.5$  & $3.95^{+0.17}_{-0.17}$  &  $1.47^{+0.14}_{-0.14}$ \\
$1.5 \le z_{\rm ph} < 2.1$  & $3.44^{+0.27}_{-0.27}$   &  $1.42^{+0.20}_{-0.20}$ \\
$z_{\rm ph} \ge 2.1$  & $4.77^{+0.26}_{-0.26}$   &  $1.40^{+0.17}_{-0.17}$\\
\enddata
\end{deluxetable}

\begin{deluxetable}{ccccc}
\tabletypesize{}
\tablecolumns{5}
\tablewidth{0pt}
\tablecaption{Effective halo masses, $M_{\rm H}$, and SFRs  derived from the effective linear bias parameters determined using jointly the reconstructed galaxy auto- and cross-spectra in the different redshift intervals. A Chabrier IMF \citep{chabrier03} was adopted to evaluate the SFR.\label{SFR_results}}
\tablehead{
\colhead{Bin} & \colhead{$b$} & \colhead{$\log{M_{\rm H}/M_{\odot}}$} &
\colhead{$\log$ SFR [$M_{\odot}$ yr$^{-1}$]}}
\startdata
$z_{\rm ph} \ge 1.5$  & $3.38^{+0.16}_{-0.16}$  &  $12.9 \pm
0.1$ & $2.6 \pm 0.2$\\
$1.5 \le z_{\rm ph} < 2.1$  & $2.59^{+0.28}_{-0.29}$   &  $12.7
\pm 0.2$  & $2.4\pm 0.2$ \\
$z_{\rm ph} \ge 2.1$  & $4.51^{+0.24}_{-0.25}$   &  $13.1 \pm
0.1$ & $2.8\pm 0.2$ \\
\enddata
\end{deluxetable}

%%%%%%%%%%%%%
%%        RESULTS      %%
%%%%%%%%%%%%%

\section{Summary and conclusions}
\label{sec:conclusions}

We have updated our previous analysis of the cross-correlation between the matter over-densities traced by the H-ATLAS galaxies and the CMB lensing maps reconstructed by the {\it Planck} collaboration. Using the new {\it Planck} lensing map we confirm the detection of the cross-correlation with a total significance now increased to 22$\,\sigma$, despite of the small area covered by the H-ATLAS survey (about $\sim 1.3\%$ of the sky) and the \emph{Planck} lensing reconstruction noise level. The improvement is due to the higher signal-to-noise ratio of {\it Planck} maps.

This result was shown to be stable against changes in the mask adopted for the survey and for different galaxy selections. A considerable effort was spent in modeling the redshift distribution, $dN/dz$, of the selected galaxies. This is a highly non-trivial task due to the large  uncertainties in the photometric redshift estimates. We have applied a Bayesian approach to derive the redshift distribution given the photo-$z$ cuts, $z_{\rm ph}$, and the r.m.s. error on $z_{\rm ph}$.

As a first step towards the investigation of the way the dark matter distribution is traced by
galaxies we have divided our galaxy sample  ($z_{\rm ph} \ge 1.5$) into two redshift intervals, $1.5 \le z_{\rm ph} < 2.1$ and $z_{\rm ph} \ge 2.1$, containing similar numbers of sources  and thus similar shot-noise levels.

A joint analysis of the cross-spectrum and of the auto-spectrum of the galaxy density contrast yielded, for the full $z_{\rm ph}\ge 1.5$ sample, a bias parameter of $b = 3.54^{+0.15}_{-0.14}$. This value differs from the one found in B15 ($b=2.80^{+0.12}_{-0.11}$) because of the different modeling of the redshift distribution, $dN/dz$: when the analysis is performed adopting the same $dN/dz$ as B15 we recover a value of $b$ very close to theirs.

On the other hand, we still find the cross-correlation amplitude to be higher than expected in the standard $\Lambda$CDM model although by a slightly smaller factor: $A=1.45^{+0.14}_{-0.13}$ against $A=1.62 \pm 0.16$, for the full galaxy sample ($z_{\rm ph}\ge 1.5$). A similar excess amplitude is found for both redshift intervals, although it is slightly larger for the lower-$z$ interval, which may be more liable to the effect of the redshift--dust temperature degeneracy, hence more affected by large failures of photometric redshift estimates. We have $A=1.48^{+0.20}_{-0.19}$ for the lower $z$ interval against $A=1.37 \pm 0.16$ for the higher $z$ one. Larger uncertainties in $z_{\rm ph}$ may be responsible, at least in part, also for the lack of power in the lowest multipole bin of the auto-power spectrum of galaxies in the lower redshift interval. However, reassuringly, the measured cross-correlation of positions of galaxies in the two redshift intervals is in good agreement with the expectations given the overlap of the estimated redshift distributions due to errors in the estimated redshifts. It is thus unlikely that the two redshift distributions are badly off.

We have also tested the dependence of the results on the assumed SED (used to estimate the redshift distribution) by repeating the full analysis using the \cite{pearson13} SED. The deviation from the expected value, $A=1$, of the cross-correlation amplitude recurs, although at a somewhat lower significance level ($\simeq 2.5\,\sigma$ instead of $\simeq 3.5\,\sigma$). However this happens at the cost of increasing the bias parameter for the higher redshift interval to values substantially higher than those given by independent estimates.

The resulting values of $A$ are found to be only marginally affected by having ignored the effect of non-linearities in the galaxy distribution and of variations of the bias parameter within each redshift interval, as well as by different choices of the background cosmological parameters.

The data indicate a significant evolution with redshift of the effective bias parameter: for our baseline redshift distributions we get $b = 2.89 \pm 0.23$ and $b=4.75^{+0.24}_{-0.25}$ for the lower- and the higher-$z$ interval, respectively. The increase of $b$ reflects a slight increase of the effective halo mass, from $\log (M_H/M_\odot)=12.7\pm 0.2$ to $\log (M_H/M_\odot)=13.1\pm 0.1$. Interestingly, the evolution of $b$ is consistent with that of the luminosity weighted bias factor yielded, as a function of $M_H$ and $z$, by the standard linear bias model. According to the SFR--$M_H$ relationships derived by \citet{aversa15}, the typical SFRs associated to these halo masses are $\log(\hbox{SFR}/M_{\odot}\,\hbox{yr}^{-1})=2.4\pm 0.2$ and $2.8\pm 0.2$, respectively.

If residual systematics in both lensing data and source selection is sub-dominant, then one would conjecture that the selected objects trace more lensing power than the bias would represent, in order to achieve a cross-correlation amplitude closer to 1.

An amplitude of the cross-correlation signal different from unity has been recently reported by the Dark Energy Survey (DES) collaboration \citep{desxc:2015} who measured the cross-correlation between the galaxy density in their Science Verification data and the CMB lensing maps provided by the {\it Planck} satellite and by the South Pole Telescope (SPT). They however found $A< 1$, but for a galaxy sample at lower
(photometric) redshifts than our sample. So, their result is not necessarily conflicting with ours, especially taking into account that they found $A$ to be increasing with redshift. Another hint of tension between $\Lambda$CDM predictions and observations has been reported by \cite{pullen15}, where the authors correlated the \emph{Planck} CMB lensing map with the Sloan Digital Sky Survey III (SDSS-III) CMASS galaxy sample at $z =0.57$, finding a tension with general relativity predictions at a $2.6\sigma$ level. In another paper, \cite{omori15} compare the linear galaxy bias inferred from measurements of the \emph{Planck} CMB lensing - CFHTLens galaxy density cross-power spectrum and the galaxy auto-power spectrum, reporting significant differences between the values found for 2013 and 2015 \emph{Planck} releases. This case has been further investigated by exploiting the analysis scheme developed in B15 by \cite{kuntz15}, where the author partially confirms the \cite{omori15} results, finding different cross-correlation amplitude values between the two \emph{Planck} releases.

The CMB lensing tomography is at an early stage of development. Higher signal-to-noise ratios will be reached due to the augmented sensitivity of both galaxy surveys, such as DES, Euclid, LSST, DESI, and of CMB lensing experiments, such as AdvACT \citep{calabrese14} or the new phase of the POLARBEAR experiment, the Simons Array \citep{polarbear14}. In the near future, the LSS will be mapped at different wavelengths out to high redshifts, enabling the comparison with the results presented in this and other works, the comprehension of the interplay between uncertainties in datasets and astrophysical modeling of sources, as well as the constraining power on both astrophysics and cosmology of cross-correlation studies.

\acknowledgments
We thank the anonymous referee for insightful comments that helped us improve the paper. F.B. would like to acknowledge David Clements, Stephen Feeney, and Andrew Jaffe for many stimulating discussions and warmly thanks the Imperial Centre for Inference and Cosmology (ICIC) for hosting him during his Erasmus Project where this work was initiated. A.L. thanks SISSA for warm hospitality. Work supported in part by INAF PRIN 2012/2013 ''Looking into the dust-obscured phase of galaxy formation through cosmic zoom lenses in the Herschel Astrophysical Terahertz Large Area Survey'' and by ASI/INAF agreement 2014-024-R.0. F.B., M.C., and C.B. acknowledge partial support from the INFN-INDARK initiative and J.G.N.  from the Spanish MINECO for a ``Ramon y Cajal''
fellowship. In this paper we made use of \lstinline!CAMB!,  \lstinline!HEALPix!,  \lstinline!healpy!, and \lstinline!emcee! packages and of the \emph{Planck} Legacy Archive (PLA).

%%%%%%%%%%%%%
%%  BIBLIO		  %%
%%%%%%%%%%%%%

\end{document}